\documentclass[a4paper,11pt]{article}
\def\bea {\begin{eqnarray}}
\def\eea {\end{eqnarray}}

\pdfoutput=1 

\usepackage{jheppub} 

\usepackage[T1]{fontenc} 

\title{\boldmath Order-by-order Anisotropic Transport Coefficients of a Magnetised Fluid: a Chapman-Enskog Approach}

\author[a]{Utsab Gangopadhyaya}
\author[a]{and Victor Roy}

\affiliation[a]{National Institute of Science Education and Research, An OCC of Homi Bhabha National Institute, 752050 Bhubaneswar, India.}

\emailAdd{utsabgang@niser.ac.in}
\emailAdd{victor@niser.ac.in}

\abstract{We derive the first and second-order expressions for the shear, the bulk viscosity, and the thermal conductivity of a relativistic 
hot boson gas in a magnetic field using the relativistic kinetic theory within the Chapman-Enskog method. The order-by-order off-equilibrium distribution function is obtained in terms of the associate Laguerre polynomial with magnetic field-dependent coefficients using the relativistic Boltzmann-Uehling-Uhlenbeck transport equation. The order-by-order anisotropic transport coefficients are evaluated in powers of the dimensionless ratio of kinetic energy to the fluid temperature for finite magnetic fields. In a magnetic field, the shear viscosity (in all order) splits into five different coefficients. Four of them show a magnetic field dependence as seen in a previous study \cite{Ashutosh1} using the relaxation time approximation for the collision kernel. On the other hand, bulk viscosity, which splits into three components (in all order), is independent of the magnetic field. The thermal conductivity shows a similar splitting but is field-dependent. The difference in the first and second-order results are prominent for the thermal conductivities than the shear viscosity; moreover, the difference in the two results is most evident at low temperatures. The first and second-order results seem to converge rapidly for high temperatures.}

\begin{document} 
\maketitle
\flushbottom

\section{Introduction} 

In linear response theory, the thermodynamic fluxes (such as momentum flow, heat flow, etc.) are assumed to be proportional to the 
thermodynamic forces ( expressed in terms of the gradients of velocities, temperature, etc.), the proportionality constants are called
transport coefficients, a few examples include shear, bulk viscosities, thermal conductivity, etc. The study of these transport coefficients for 
hot and dense nuclear matter, a.k.a. Quark-Gluon Plasma (QGP) produced at Relativistic Heavy Ion Collider (RHIC) and the Large Hadron Collider (LHC) experiments, is one of the primary goals since it was found that the QGP has smallest specific shear viscosity.
Also, nuclear collisions at RHIC and LHC experiments produce intense transient electromagnetic fields due to the
relativistic velocity of the protons inside the colliding nucleus. The effect of these transient electromagnetic fields
on the hot and dense QGP and the subsequent low-temperature phase of hadronic resonance gas is an active 
area of research ~\cite{Deng:2012pc,Guo:2017jxs,Satow,Skokov,Panda:2021pvq,Panda:2020zhr,Biswas:2020rps,Mohanty:2018eja,Most:2021uck,Most:2021rhr,Roy:2015coa,Roy:2015kma,Pu:2016ayh,Roy:2017yvg,Dash:2017rhg,Denicol:2019iyh,Denicol:2018rbw,Inghirami:2019mkc,Inghirami:2016iru,Ghosh:2022xtv, Bandyopadhyay:2017cle,Karmakar:2020mnj,Karmakar:2018aig,Shokri:2018qcu,Singh:2022ltu,Bhadury:2022qxd,Pang:2016yuh,Kurian:2020kct,Kurian:2019nna,K:2021sct,Chatterjee:2017ahy,Chatterjee:2018lsx,Ghosh:2020wqx,Hattori:2022hyo}. Magnetic fields in the mid-central and peripheral heavy-ion collisions predominantly point perpendicular to the participant plane. This breaks the isotropic symmetry in the momentum space in the transverse plane. While subjected to these directed magnetic fields, the quarks in the QGP phase and the charged hadrons in the HRG phase may give rise to anisotropic transport phenomena~\cite{A.Das1,A.Das2,A.Das3,Huang:2009ue,Hattori:2017qih} and associated anisotropic transport coefficients. It is known that some of these new transport coefficients do not contribute to the entropy generation  and hence are non-dissipative ~\cite{Hernandez:2017mch,Hess}. These anisotropic transport coefficients were studied in Ref.~\cite{Ashutosh1} 
using relativistic Boltzmann equation in the RTA approximation. Here we use the modified Chapman-Enskog (CE) method~\cite{De.Groot,De.Groot2,Mitra:2013gya,Mitra:2014dia} for solving the relativistic 
Boltzmann-Uehling-Uhlenbeck~(BUU) transport equation in the presence of magnetic fields and calculate anisotropic transport coefficients.
In the CE method, we expand the collisions integral in terms of a complete set of orthogonal 
polynomials with magnetic field-dependent coefficients. This method can calculate the transport coefficients to a desired degree 
of accuracy in powers of dimensionless kinetic energy (normalized by temperature) compared to the previous methods; in this approach, the scaled kinetic energy acts as a smallness parameter for the expansion. 
This paper is organised as follows: In Sec.~\eqref{sec:Boltzmann} we discuss the Boltzmann equation and the specific form of the linearised collision
integral that is expanded in terms of the orthogonal polynomial. We also give analytical expressions for the magnetic field-dependent first and second-order transport coefficients. In Sec.~\eqref{sec:numeric} we discuss the numerical results, including the temperature and magnetic field dependence of
anisotropic shear, bulk viscosities, and heat conductivity. Finally, we summarise this work in Sec.~\eqref{sec:summ}. Some detailed derivations of the useful formulas are given in App.~\eqref{projection}-\eqref {app:Other}.
Throughout the manuscript we use natural unit $\bar{h}=c=k_B=1$ and mostly negative metric  $g^{\mu\nu}=diag(+1,-1,-1,-1)$.

\section{Linearised Boltzmann Equation and Transport Coefficients}
\label{sec:Boltzmann}

In this work, we primarily follow the formalism of Ref.~\cite{Davesne}, which considers Bose enhancement factors in the collision
integral of the Boltzmann equation, but here we also consider the presence of finite magnetic fields. In this case,
the Boltzmann equation contains a force term due to the magnetic field and has the following form:
\bea
\label{eq:Boltzmann}
p^{\mu}\partial_{\mu}f+qF^{\mu\nu}p_{\nu}\frac{\partial f}{\partial p^{\mu}}=C(f),
\eea
where, the collision integral $C(f)=\int d\Gamma_{k}d\Gamma_{p'}d\Gamma_{k'}\{f_{p'}f_{k'}(1+ A_{0}f_{p})(1+A_{0} f_{k})-f_{p}f_{k}(1+ A_{0}f_{p'})(1+ A_{0}f_{k'})\}W$  and the transition rate $W=\frac{s}{2}\frac{d\sigma}{d\Omega}(2\pi)^{6}\delta^{4}(p+k-p'-k'), d\Gamma_{q}=\frac{d^{3}q} {(2\pi)^3q_{0}},~ q_{0}=\sqrt{\vec{q}^{2}+m^{2}}$. In the present work we consider only pion gas, the relevant cross sections for $\pi\text{-}\pi$ reactions 
can be found in Ref.~\cite{Bertsch,Welke}, we will be using a parametrised form of the iso-spin averaged differential cross-section given by, 

\bea
\frac{d\sigma(s)}{d\Omega}=\frac{4}{q_{cm}^{2}}\left[\frac{1}{9}\sin^{2}\delta^{0}_{0}+\frac{1}{3}9\sin^{2}\delta^{1}_{1}\cos^{2}\theta +\frac{5}{9}\sin^{2}\delta^{2}_{0}\right],
\eea 
where,
\bea
\delta^{0}_{0}&=&\frac{\pi}{2}+\tan^{-1}\left(\frac{E-m_{\sigma}}{\Gamma_{\sigma}/2}\right),    \nonumber\\
\delta^{1}_{1}&=&\frac{\pi}{2}+\tan^{-1}\left(\frac{E-m_{\rho}}{\Gamma_{\rho}/2}\right),   \nonumber\\
\delta^{2}_{0}&=&-0.12q_{cm}/m_{\pi}.
\eea 

The widths, $\Gamma_{\sigma}=2.06q_{cm}$ and $\Gamma_{\rho}=0.095q_{cm}\left(\frac{q_{cm}/m_{\pi}}{1+\left(q_{cm}/m_{\pi}\right)^{2}}\right)^{2}$ with $m_{\sigma}=5.8m_{\pi}$ and $m_{\rho}=5.53m_{\pi}$. Here $\sqrt{s}= E$ is the  collision energy in the center of mass frame; $q_{cm}=\sqrt{\left(\frac{E}{2}\right)^2-m_{\pi}^{2}}$.

The collision integral vanishes  for the local-equilibrium distribution function $f_{0}$, which for our case has the following form 
$f_{0}=A_{0}^{-1}\left[\exp\left(\frac{p^{\mu}U_{\mu}-\mu}{T}\right) - 1\right]^{-1}$, here $T$ is the temperature and $U_{\mu}$ is the time-like
four velocity, $\mu$ is the chemical potential. If the deviation from local equilibrium is small, we can assume the off-equilibrium
single particle distribution can be expanded in terms of a small parameter $\epsilon$ as:
\begin{equation}
f(x,p)=f^{0}(x,p)+\epsilon f^{1}(x,p)+\epsilon^{2}f^{2}(x,p)+...~.
\label{totalF}
\end{equation}
To be specific, we use the Knudsen number (the ratio of mean free path to system size) as $\epsilon$ and use the Chapman-Enskog
method to calculate the off-equilibrium part of the distribution function. In equilibrium, the energy momentum tensor is defined as the 
second moment of $f_{0}$, i.e., $T^{\mu\nu}_{0}=\int d\Gamma p^{\mu}p^{\nu}f_{0}$, the equilibrium energy density and the pressure can
be calculated as $\varepsilon_{0}=U_{\mu}U_{\nu}T^{\mu\nu}_{0}$, and $p=-\frac{1}{3}\Delta_{\mu\nu}T^{\mu\nu}$; here $\Delta_{\mu\nu}=(g^{\mu\nu}-U^{\mu}U^{\nu})$  is the projector operator orthogonal to $U^{\mu}$. From the above definitions, one can identify the parameter 
$T$ in $f_0$ as the equilibrium local thermodynamic temperature. When considering an out-of-equilibrium system, identifying the temperature and the fluid four velocities is somewhat ad-hoc and require extra constraints to define them. This is usually done by choosing a specific frame; two widely used frames are Landau-Lifshitz(LL) and Eckart(EK). As per the EK definition, $U_{\mu}$ is parallel to the particle four flow $N^{\mu}=nU^{\mu}=\int d\Gamma p^{\mu}f$; hence $N^{\mu}$ does not contain any dissipative corrections. Here we notice that the use of the EK frame demands that there must be some conserved charges $n$, for our study we consider number of pions.
With this choice of $U_{\mu}$ the energy-momentum tensor can be written as $T^{\mu\nu}=T^{\mu\nu}_{0}+\delta T^{\mu\nu}$, where 
$T^{\mu\nu}_{0}=\varepsilon U^{\mu}U^{\nu}+p\Delta^{\mu\nu}$ and $\delta T^{\mu\nu}= \Pi^{\mu\nu}+\left[\left(I^{\mu}+h\Delta^{\mu\sigma}N_{\sigma}\right)U^{\nu}+\left(I^{\nu}+h\Delta^{\nu\sigma}N_{\sigma}\right)U^{\mu}\right]$, where $ \Pi^{\mu\nu}$
is the viscous stress tensor and $I^{\mu}=(U_{\nu}T^{\nu\rho}\Delta_{\rho}^{\mu}-h\Delta^{\mu\sigma}N_{\sigma})$; $h$ is the enthalpy. The viscous part is further split into the traceless shear part $\pi^{\mu\nu}$, and non-zero trace bulk part $\Pi$ as $\Pi^{\mu\nu}=\pi^{\mu\nu}+\Pi\Delta^{\mu\nu}$. In Eckart 
frame $h\Delta^{\mu\sigma}N_{\sigma}$ vanishes in the expression for $I^{\mu}$. While expressing $f(x,p)$ as in Eq.\eqref{totalF}
$\mu$, $T$, and $U^{\mu}$ are taken to be as parameters; with the further assumptions that $n=\int d\Gamma U_{\mu}p^{\mu} f=\int d\Gamma U_{\mu}p^{\mu}f_{0}=n_{0}$ and $\varepsilon=\int d\Gamma \left(U_{\mu}p^{\mu}\right)^2 f=\int d\Gamma \left(U_{\mu}p^{\mu}\right)^2f_{0}=\varepsilon_{0}$ we can assign $T$ and $\mu$ as temperature and the chemical potential in the off-equilibrium case. This can be clearly
seen from the fact that the number, energy density, and pressure can be expressed as:
\begin{eqnarray}
	n&=& \frac{z^{2}T^{3}A_{0}^{-1}}{2\pi^{2}} S_{2}^{-1}, \\
	\varepsilon&=&\frac{z^{2}T^{4}A_{0}^{-1}}{2\pi^{2}}\left[zS^{-1}_{3}-S^{-2}_{2}\right], \\
	p&=&nT\frac{S_{2}^{-2}}{S_{2}^{-1}}.
\end{eqnarray}
Here $z=m/ T$, $S_{n}^{\alpha}=\Sigma_{k=1}^{\infty} e^{k \mu / T} k^{\alpha} K_{n}(k z)$, where $K_{n}(x)$ is the modified Bessel
function of the second kind.
With this brief introduction, let us discuss the linearised form of the Boltzmann equation used to calculate
the anisotropic transport coefficients in magnetic fields.
Introducing $\epsilon$ in the space-time derivative term of the Boltzmann equation Eq.~\eqref{eq:Boltzmann}, using
Eq.~\eqref{totalF} for the off-equilibrium distribution function we get 

\bea
p^{\mu}\epsilon\partial_{\mu}f+qF^{\mu\nu}p_{\nu}\frac{\partial f}{\partial p^{\mu}}&=&C(f),\nonumber\\
p^{\mu}\epsilon \partial_{\mu}\left(f^{0}+\epsilon f^{1}+...\right) +qF^{\mu\nu}p_{\nu}\frac{\partial }{\partial p^{\mu}}\left(f^{0}+\epsilon f^{1}+...\right)&=&C(f).
\eea 
 The book keeping parameter  $\epsilon$ has been introduced in the LHS (also in Eq.\eqref{totalF})  to count the order of space derivatives 
of the macroscopic parameters that determines the off-equilibrium distribution function. Now , matching the coefficients of same power in $\epsilon$ we get a hirearchy of equations; to the first order we have:
\bea
\label{eq:BUU}
p^{\mu}\partial_{\mu}f^{0}=-\mathcal{L}(f^{1})-qF^{\mu\nu}p_{\nu}\frac{\partial f^{1}}{\partial p^{\mu}},
\eea 
where  
\bea
\label{eq:Collterm}
C(f^{1})=-\mathcal{L}(f^{1})=-\int d\Gamma_{k}d\Gamma_{p'}d\Gamma_{k'}f^{0}(x,p)f^{0}(x,k)\left(1+A_{0}f^{0}(x,p')\right)\nonumber\\
\times\left(1+A_{0}f^{0}(x,k')\right)\left(\phi(x,p)+\phi(x,k)-\phi(x,p')-\phi(x,k')\right)W
\eea 
and  $f^{1}=f^{0}(1\pm f^{0})\phi$. Here we consider only Bosonic system (appropriate terms in the collision kernel is taken into account) which leads to the BUU equation Eq.~\eqref{eq:BUU}.
Now expanding the left hand side of the BUU equation,
\bea
\label{eq:LHSBUU}
\nonumber
\frac{f^{0}\left(1+A_{0}f^{0}\right)}{T}  \bigg[ Q\nabla_{\sigma}U^{\sigma}-\frac{1}{2}p^{\mu}p^{\nu}\left(\nabla_{\mu}U_{\nu}+\nabla_{\nu}U_{\mu}-\frac{2}{3}\triangle_{\mu\nu}\nabla_{\sigma}U^{\sigma}\right)    \nonumber\\
+\left(p\cdot U-h\right)p^{\mu}\triangle_{\mu}^{\sigma}T^{-1}\left(\partial_{\sigma}T-TDU_{\sigma}\right)\bigg],
\eea 
Where,
\bea	
Q=-\frac{m^{2}}{3}+\left(p\cdot U\right)^{2}\left(\frac{4}{3}-\gamma'\right)+\left[\left(\gamma''-1\right)h-\gamma'''T\right]\left(p\cdot U\right),   \label{Q}
\eea
and the terms $\gamma'$, $\gamma''$ and $\gamma'''$ are expressed as,
\bea
\gamma'&=&\frac{\left(S^{0}_{2}/S^{-1}_{2}\right)^{2}-\left(S^{0}_{3}/S^{-1}_{2}\right)^{2}+4z^{-1}S^{0}_{2}S^{-1}_{3}/\left(S^{-1}_{2}\right)^{2}+z^{-1}S^{0}_{3}/S^{-1}_{2}}{\left(S^{0}_{2}/S^{-1}_{2}\right)^{2}-\left(S^{0}_{3}/S^{-1}_{2}\right)^{2}+3z^{-1}S^{0}_{2}S^{-1}_{3}/\left(S^{-1}_{2}\right)^{2}+2z^{-1}S^{0}_{3}/S^{-1}_{2}-z^{-2}}, \\
\gamma''&=&1+\frac{z^{-2}}{\left(S^{0}_{2}/S^{-1}_{2}\right)^{2}-\left(S^{0}_{3}/S^{-1}_{2}\right)^{2}+3z^{-1}S^{0}_{2}S^{-1}_{3}/\left(S^{-1}_{2}\right)^{2}+2z^{-1}S^{0}_{3}/S^{-1}_{2}-z^{-2}},~~~~~ \\
\gamma'''&=&\frac{S^{0}_{2}/S^{-1}_{2}+5z^{-1}S^{-1}_{3}/S^{-1}_{2}-S^{0}_{3}S^{-1}_{3}/\left(S^{-1}_{2}\right)^{2}}{\left(S^{0}_{2}/S^{-1}_{2}\right)^{2}-\left(S^{0}_{3}/S^{-1}_{2}\right)^{2}+3z^{-1}S^{0}_{2}S^{-1}_{3}/\left(S^{-1}_{2}\right)^{2}+2z^{-1}S^{0}_{3}/S^{-1}_{2}-z^{-2}}. 
\eea 
Using the notation $f^{0}(x,p)\equiv f^{0}_p$, the right hand side of Eq.\eqref{eq:BUU}  can be written as :
\bea
\nonumber
-\int d\Gamma_{k}d\Gamma_{p'}d\Gamma_{k'}\big[f^{0}_{p}f^{0}_{k}\left(1+A_{0}f^{0}_{p'}\right)\left(1+A_{0}f^{0}_{k'}\right) \left(\phi_{p}+\phi_{k}-\phi_{p'}-\phi_{k'}\right)W\big]  \nonumber\\
-qF^{\mu\nu}p_{\nu}\frac{\partial}{\partial p_{\mu}}\left[f^{0}\left(1+A_{0}f^{0}\right)\phi\right].
\eea 
In this work we consider only magnetic field, in this case the Maxwell stress tensor $F^{\mu\nu}=-Bb^{\mu\nu}$ where $b^{\mu\nu}=\epsilon^{\mu\nu\rho\alpha}\frac{B_{\rho}}{B}U_{\alpha}$.
Here $B^{\rho}$ is the magnetic four vector and $B=\sqrt{B^{\rho}B_{\rho}}$ is the magnitude of the magnetic field in the rest frame. 
The last term on the RHS gives :
\begin{equation}
\label{eq:MDBUU}
qBb^{\mu\nu}p_{\nu}f^{0}\left(1+A_{0}f^{0}\right)\frac{\partial\phi_{p}}{\partial p^{\mu}}.
\end{equation}
Where we have used the relation $\epsilon^{\mu\nu\rho\alpha}B_{\rho}U_{\alpha}p_{\nu}U_{\mu}=0$ in deriving the above expression.
Our job is to find the expression for $\phi$ which will enable us to calculate the transport coefficients discussed in the next few sections.

\subsection{Shear Viscosity}
\label{sec:shear}
In this section we derive the expression for shear viscous coefficients. Using Eq.~\eqref{eq:BUU}, Eq.~\eqref{eq:LHSBUU}, and Eq.~\eqref{eq:MDBUU} we have
\bea
\label{eq:ShearBUU}
\frac{f^{0}\left(1+A_{0}f^{0}\right)}{2T}p^{\mu}p^{\nu}\left(\nabla_{\mu}U_{\nu}+\nabla_{\nu}U_{\mu}-\frac{2}{3}\triangle_{\mu\nu}\nabla_{\sigma}U^{\sigma}\right)=\mathcal{L}(\phi)-f^{0}\left(1+A_{0}f^{0}\right)\mathcal{F}^{\mu}\frac{\partial \phi_{p}}{\partial p^{\mu}}. ~~~~~~
\eea 
Here we only consider the traceless symmetric part of the velocity gradient, and $\mathcal{F}^{\mu}=qBb^{\mu\nu}p_{\nu}$. 
Since magnetic field breaks the isotropy of the system the most general form of $\phi$ for shear viscosity only can be written as Ref.~\cite{Ashutosh1}: 
\bea
\label{eq:phi}
\phi = \sum_{n=0}^{4}\mathcal{X}_{n}C_{(n)\mu\nu\alpha\beta}~\langle p^{\mu}p^{\nu}\rangle V^{\alpha\beta},
\eea 
where,
\bea
V^{\eta\delta} =\frac{1}{2}\left[\triangle^{\eta\lambda}\triangle^{\delta\kappa}+\triangle^{\eta\kappa}\triangle^{\delta\lambda}\right]\partial_{\lambda}U_{\kappa};~~~~\mathcal{X}_{n}=\mathcal{X}_{n}(B,\tau');\\
\tau'=\tau-z;~~~~~~\tau=\frac{p\cdot U}{T};~~~~~~z=\frac{m}{T}.~~~~~~~~~
\eea 
The  expansion coefficient $\mathcal{X}_{n}\left(B,\tau'\right)$ are expressed in terms of the orthogonal Laguerre polynomial of order $\frac{5}{2}$ and degree $m$ as Ref.~\cite{Davesne},
\bea
\label{eq:Xn}
\mathcal{X}_{n}\left(B,\tau'\right)=\sum_{m=0}^{\infty}c^{\left(n\right)}_{m}\left(B\right)L_{m}^{\frac{5}{2}}\left(\tau'\right).   \label{Series_1}
\eea
$C_{(n)\mu\nu\alpha\beta}$ are the projection tensors, whose properties have been discussed in details in App.~(\ref{projection}) and in Ref.~\cite{Hess,Ashutosh1}.
The particular form of $\phi$ was chosen so that we have a similar tensorial structure on both sides of Eq.~\eqref{eq:ShearBUU}. Once $\phi$ is
known, we can evaluate the shear stress tensor as: 
\bea
\pi^{\mu\nu}=\int d\Gamma \langle p^{\mu}p^{\nu}\rangle \delta f=\int d\Gamma \langle p^{\mu}p^{\nu}\rangle f^{0}\left(1+A_{0}f^{0}\right)\sum_{n=0}^{4}\mathcal{X}_{n}C_{\left(n\right)\alpha\beta\eta\delta}p^{\alpha}p^{\beta}V^{\eta\delta}.   
\eea 

Here $\langle A^{\mu}B^{\nu} \rangle=\Delta^{\mu\nu}_{\alpha\beta}A^{\alpha}B^{\beta}$; $\Delta^{\mu\nu}_{\alpha\beta}= \frac{1}{2}\left[\triangle^{\mu}_{\alpha}\triangle^{\nu}_{\beta}+\triangle^{\mu}_{\beta}\triangle^{\nu}_{\alpha}-\frac{2}{3}\triangle^{\mu\nu}\triangle_{\alpha\beta}\right] $.
For a system with anisotropy, the fourth-rank shear-viscous coefficient $\eta^{\mu\nu\psi\theta}$ is defined as:
\bea
\label{eq:shearfourth}
\pi^{\mu\nu}=\eta^{\mu\nu\psi\theta}\langle\partial_{\psi}U_{\theta} \rangle=\sum_{n=0}^{4}\eta_{n}C_{\left(n\right)}^{\mu\nu\psi\theta}\langle\partial_{\psi}U_{\theta} \rangle, 
\eea 
where $\eta_{n}$ are the unknown coefficients to be determined by using appropriate tensor contractions defined in App.~(\ref{projection}). 
The final expressions are:
\bea
\eta_{\parallel}=\eta_{0}=\frac{\rho T^{2}}{2}\sum_{n=0}^{\infty}c_{n}^{\left(0\right)}\gamma_{n}^{\left(0\right)};~~~~~~~~~~~~~~~~~   \label{eta0}    \\
\eta_{\perp}=\eta_{1}=\frac{\rho T^{2}}{4}\sum_{n=0}^{\infty}c_{n}^{\left(1\right)}\gamma_{n}^{\left( 1\right)};~~~~~~\eta_{\times}= \eta_{2}=\frac{\rho T^{2}}{4}\sum_{n=0}^{\infty}c_{n}^{\left(2\right)}\gamma_{n}^{\left( 1\right)};   \label{eta12}   \\
\eta'_{\perp}=\eta_{3}=\frac{\rho T^{2}}{4}\sum_{n=0}^{\infty}c_{n}^{\left(3\right)}\gamma_{n}^{\left( 3\right)};~~~~~~
\eta'_{\times}=\eta_{4}=\frac{\rho T^{2}}{4}\sum_{n=0}^{\infty}c_{n}^{\left(4\right)}\gamma_{n}^{\left( 3\right)}.   \label{eta34}
\eea 
Here $\gamma^{\left(n\right)}_{j}=\frac{1}{\rho T^{2}}\int d\Gamma_{p}f^{0}\left(1+A_{0}f^{0}\right)\langle p_{\alpha}p_{\beta}\rangle \langle p_{\eta}p_{\delta}\rangle L^{\frac{5}{2}}_{j}\left(\tau_{p}\right)C_{\left(n\right)}^{\alpha\beta\eta\delta}$ ; for brevity, henceforth, we do not explicitly write the $B$ dependence in $c_{n}^{(j)}$.
The constraining equations for the coefficients $c_{n}^{(j)}$ are obtained by using Eq.~\eqref{eq:phi} in the BUU equation Eq.~\eqref{eq:ShearBUU}: 
\bea
\frac{1}{2\rho T}\gamma^{\left(0\right)}_{j}=&\sum_{n=0}^{\infty}c_{n}^{\left(0\right)}c_{nj}; ~~~~~~~~~~~~~~~~~~~~~~~~~~~~~~~~~~~\label{c0_relation}    \\
\frac{1}{2\rho T}\gamma^{\left(1\right)}_{j}=\sum_{n=0}^{\infty}c_{n}^{\left(1\right)}d_{nj}+\sum_{n=0}^{\infty}c_{n}^{\left(2\right)}\xi_{nj}^{\left(1\right)};&~~~
\frac{1}{2\rho T}\gamma^{\left(2\right)}_{j}=\sum_{n=0}^{\infty}c_{n}^{\left(1\right)}\xi_{nj}^{\left(1\right)}-\sum_{n=0}^{\infty}c_{n}^{\left(2\right)}d_{nj};  ~~ \label{c1_relation}\\
\frac{1}{2\rho T}\gamma^{\left(3\right)}_{j}=\sum_{n=0}^{\infty}c_{n}^{\left(3\right)}l_{nj}+2\sum_{n=0}^{\infty}c_{n}^{\left(4\right)}\xi_{nj}^{\left(3\right)};&~~~
\frac{1}{2\rho T}\gamma^{\left(4\right)}_{j}=2\sum_{n=0}^{\infty}c_{n}^{\left(3\right)}\xi_{nj}^{\left(3\right)}-\sum_{n=0}^{\infty}c_{n}^{\left(4\right)}l_{nj}.~~\label{c2_relation}
\eea 

One can calculate the transport coefficients to a desired degree by truncating the series Eq.~(\ref{Series_1}) to the desired order. The expression for the first-order shear viscosity are as follows:
\bea
\label{eq:eta01}
\left[\eta_{\parallel}\right]_{1}=\left[\eta_{0}\right]_{1}&=&\frac{T}{4}\frac{\left(\gamma_{0}^{\left(0\right)}\right)^{2}}{c_{00}}=\frac{T}{10}\frac{\gamma_{0}^{2}}{c'_{00}},  \\
\left[\eta_{\perp}\right]_{1}=\left[\eta_{1}\right]_{1}&=&\frac{T}{8}\left[\frac{\left(\gamma_{0}^{\left(1\right)}\right)^{2}d_{00}}{\left(d_{00}\right)^{2}+\left(\xi_{00}^{\left(1\right)}\right)^{2}}\right],    \\
\left[\eta_{\times}\right]_{1}=\left[\eta_{2}\right]_{1}&=&\frac{T}{8}\left[\frac{\left(\gamma_{0}^{\left(1\right)}\right)^{2}\xi^{\left(1\right)}_{00}}{\left(d_{00}\right)^{2}+\left(\xi_{00}^{\left(1\right)}\right)^{2}}\right],   \\
\left[\eta'_{\perp}\right]_{1}=\left[\eta_{3}\right]_{1}&=&\frac{T}{8}\left[\frac{\left(\gamma_{0}^{\left(3\right)}\right)^{2}l_{00}}{\left(l_{00}\right)^{2}+\left(2\xi_{00}^{\left(3\right)}\right)^{2}}\right],    \\
\label{eq:eta41}
\left[\eta'_{\times}\right]_{1}=\left[\eta_{4}\right]_{1}&=&\frac{T}{8}\left[\frac{2\left(\gamma_{0}^{\left(3\right)}\right)^{2}\xi^{\left(3\right)}_{00}}{\left(l_{00}\right)^{2}+\left(2\xi_{00}^{\left(3\right)}\right)^{2}}\right]. 
\eea 
The definitions of various symbols in the above set of equations are given in App.~(\ref{Appendix_Shear}).
We also report expression for the second-order shear viscosities and detailed calculations in App.~(\ref{Appendix_Shear})
due to length constraints.

\subsection{Bulk Viscosity}
The bulk viscosity near the crossover temperature (from the hadronic phase to the QGP phase) is conjectured to show a peak; hence
for the temperature range achieved in high energy heavy-ion collisions, bulk viscosity is an important transport coefficient to study along 
with the shear viscosity. The bulk viscosity is associated with the non-zero trace part of the dissipative correction to the energy-momentum 
tensor. In terms of $\phi$, the bulk stress can be written as:
\bea
\label{eq:genBulk}
\Pi= \zeta^{\eta\delta} \partial_{\eta}U_{\delta}=\frac{\Delta_{\mu\nu}}{3}\int d\Gamma_{p}p^{\mu}p^{\nu}f^{0}\left(1+A_{0}f^{0}\right) \phi.   \label{Bulk1}
\eea 
We can further express $\zeta^{\eta\delta} $ in terms of the projection basis as $ \zeta_{||}C_{\left(0\right)}^{\eta\delta}+\zeta_{\perp}C_{\left(1\right)}^{\eta\delta}+\zeta_{\times}C_{\left(2\right)}^{\eta\delta}$. Now, the explicit expression of $\phi$ for the bulk viscosity in presence of the magnetic field is $\phi=\sum_{n=0}^{2}\mathcal{A}_{n}C_{\left(n\right)}^{\eta\delta}\partial_{\eta}U_{\delta}$. Using this expression for $\phi$ in  Eq.~\eqref{eq:genBulk} we have 
\bea
\nonumber
\label{eq:bulk}
 \left(\zeta_{||}C_{\left(0\right)}^{\eta\delta}+\zeta_{\perp}C_{\left(1\right)}^{\eta\delta}+\zeta_{\times}C_{\left(2\right)}^{\eta\delta}\right)\partial_{\eta}U_{\delta}=\frac{\Delta_{\mu\nu}}{3}\int d\Gamma_{p}p^{\mu}p^{\nu}f^{0}\left(1+A_{0}f^{0}\right)\left[\sum_{n=0}^{2}\mathcal{A}_{n}C_{\left(n\right)}^{\eta\delta}\right]\partial_{\eta}U_{\delta}. \\
\eea
Now comparing the coefficients of linearly independent $C^{\eta\delta}_{(n)}$ on both sides of Eq.~\eqref{eq:bulk} we get,
\bea
\label{eq:zetapar}
\zeta_{||}=\frac{1}{3}\int d\Gamma_{p} f^{0}\left(1+A_{0}f^{0}\right)\triangle_{\mu\nu}p^{\mu}p^{\nu}\mathcal{A}_{0},\\
\label{eq:zetaperp}
\zeta_{\perp}=\frac{1}{3}\int d\Gamma_{p} f^{0}\left(1+A_{0}f^{0}\right)\triangle_{\mu\nu}p^{\mu}p^{\nu}\mathcal{A}_{1},\\
\label{eq:zetacross}
\zeta_{\times}=\frac{1}{3}\int d\Gamma_{p} f^{0}\left(1+A_{0}f^{0}\right)\triangle_{\mu\nu}p^{\mu}p^{\nu}\mathcal{A}_{2}.
\eea 
With the help of the following matching conditions $\int d\Gamma (U\cdot p) f^{0}\left(1+A_{0}f^{0}\right)\phi=\int d\Gamma (U\cdot p)^2 f^{0}\left(1+A_{0}f^{0}\right)\phi=0$, and using the definition of $Q$ (Eq.(\ref{Q}))
we can express  Eqs.~\eqref{eq:zetapar}, \\ \eqref{eq:zetaperp},\eqref{eq:zetacross} as

\bea
\zeta_{||}&=&\int d\Gamma_{p} f^{0}\left(1+A_{0}f^{0}\right)Q\mathcal{A}_{0},\\
\zeta_{\perp}&=&\int d\Gamma_{p} f^{0}\left(1+A_{0}f^{0}\right)Q\mathcal{A}_{1},\\
\zeta_{\times}&=&\int d\Gamma_{p} f^{0}\left(1+A_{0}f^{0}\right)Q\mathcal{A}_{2}.
\eea
To evaluate $\mathcal{A}_{(n)}$ we consider the BUU equation for the thermodynamic force $\nabla_{\alpha}U^{\alpha}$
\bea
\frac{f^{0}\left(1+A_{0}f^{0}\right)}{T}Q\nabla_{\sigma}U^{\sigma}=-\sum_{n=0}^{2}\int d\Gamma_{k}d\Gamma_{p'}d\Gamma_{k'}\Big[f^{0}_{p}f^{0}_{k}\left(1+A_{0}f^{0}_{p'}\right)\left(1+A_{0}f^{0}_{k'}\right)\nonumber\\
\times\left(\mathcal{A}_{n}^{p}+\mathcal{A}_{n}^{k}-\mathcal{A}_{n}^{p'}-\mathcal{A}_{n}^{k'}\right)WC_{\left(n\right)\eta\delta}\partial^{\eta}U^{\delta}\Big]~\nonumber\\
-qF^{\mu\nu}p_{\nu}f^{0}\left(1+A_{0}f^{0}\right)\frac{\partial}{\partial p^{\mu}}\left(\sum_{n=0}^{2}\mathcal{A}_{n}\left(B,\tau'\right)C_{\left(n\right)\eta\delta}\right)\partial^{\eta}U^{\delta}.
\eea 
The last term in the above equation vanishes because
\bea
qF^{\mu\nu}p_{\nu}\frac{\partial}{\partial p^{\mu}}\left(\sum_{n=0}^{2}\mathcal{A}_{n}\left(\tau\right)C_{\left(n\right)\eta\delta}\right)=qF^{\mu\nu}p_{\nu}\left(\sum_{n=0}^{2}\frac{\partial \mathcal{A}_{n}}{\partial \tau'}\frac{\partial\left(p\cdot U\right)}{\partial p^{\mu}}C_{\left(n\right)\eta\delta}\right)=0,
\eea 
where we have used the fact that $F^{\mu\nu}U_{\nu}=0$.
Now contracting the Boltzmann equation with $P^{\left(n\right)\eta\delta}$ (defined in App.~\ref{projection}) and making use of the properties mentioned therein we get,
\bea
\left[\mathcal{A}_{1},L_{j}^{\frac{1}{2}}\left(\tau'\right)\right]=\left[\mathcal{A}_{0},L_{j}^{\frac{1}{2}}\left(\tau'\right)\right]=\frac{m}{\rho}\alpha_{j},
\eea 
and $\mathcal{A}_{2}=0$, where $\alpha_{j}$ is a function of $z, S^{\alpha}_{n},\gamma$ and thermodynamic quantities (see Eqs.~(\ref{alpha_j})) of App.~(\ref{Appendix_Bulk}) for details). Now expanding $\mathcal{A}_{(n)}$'s using Laguerre polynomial $\mathcal{A}_{\left(n\right)}\left(B,\tau'\right)=\sum_{i=0}^{\infty}a^{\left(n\right)}_{i}\left(B\right)L^{\frac{1}{2}}_{i}\left(\tau'\right)$ we have,
\bea
\sum_{n=0}^{\infty}a^{\left(1\right)}_{n}a_{nj}=\sum_{n=0}^{\infty}a^{\left(0\right)}_{n}a_{nj}=\frac{m}{\rho}\alpha_{j}.
\eea 
Restricting the expansion of $a_{n}$ after first term we get the first-order expression for bulk viscosities,
\bea
\left[\zeta_{||}\right]_{1}=\left[\zeta_{\perp}\right]_{1}=\frac{\rho T}{m}a_{0}^{\left(2\right)}\alpha_{2}=T\frac{\alpha^{2}}{a_{22}};    ~~~ \zeta_{\times}=0. 
\eea 
We see that $\zeta$'s are independent of $B$ which corroborates the finding of Ref.~\cite{Ashutosh1,Denicol:2018rbw}. Similarly we calculate the second-order expressions for 
 $\zeta$'s by truncating the series after second-order,
\bea
\left[\zeta_{||}\right]_{2}=\left[\zeta_{\perp}\right]_{2}=T\frac{\left(a_{33}\alpha_{2}^{2}-2\alpha_{2}\alpha_{3}a_{23}+a_{22}\alpha_{3}^{2}\right)}{\left(a_{22}a_{33}-a_{23}a_{32}\right)}; ~~ \zeta_{\times}=0.
\eea 

The detailed calculation, along with the explicit expressions for various terms is given in App.~(\ref{Appendix_Bulk}).


\subsection{Thermal Conductivity}
Another quantity of interest is the thermal conductivity of hot and dense nuclear matter.
The expression for the reduced heat flow is
\begin{equation}
I^{\mu}_{q}=\left(U_{\nu}T^{\nu\sigma}-hN^{\sigma}\right)\triangle^{\mu}_{\sigma},
\end{equation}
where $h=\left(\varepsilon+p\right)/n$ is the enthalpy per particle  of the system. In linear response theory the reduced heat flow is 
proportional to the temperature-gradient and fluid acceleration $I^{\mu}_{q}=-\lambda^{\mu\nu}\triangle^{\beta}_{\nu}\left(\partial_{\beta}T-TDU_{\beta}\right)$, when expressed using the kinetic theory definition, gives: 
\bea
-\lambda^{\mu\nu}\triangle^{\beta}_{\nu}\left(\partial_{\beta}T-TDU_{\beta}\right)=\int d\Gamma_{p}~p^{\sigma}\triangle^{\mu}_{\sigma}\left(p\cdot U-h\right)f^{0}\left(1+A_{0}f^{0}\right)\phi.
\eea 
The functional form of $\phi$ in this case was chosen considering the linearised BUU equation for heat flow
\bea
\label{eq:BUUheat}
\frac{f^{0}\left(1+f^{0}\right)}{T}\left(p\cdot U-h\right)p^{\mu}\triangle_{\mu}^{\sigma}T^{-1}\left(\partial_{\sigma}T-TDU_{\sigma}\right) ~~~~~~~~~~~~~~~~~~~~~~~~~~~         \nonumber\\
=-\int d\Gamma_{k}d\Gamma_{p'}d\Gamma_{k'}~f^{0}_{p}f^{0}_{k}\left(1+A_{0}f^{0}_{p'}\right)\left(1+A_{0}f^{0}_{k'}\right)\left(\phi_{p}+\phi_{k}-\phi_{p'}-\phi_{k'}\right)W         \nonumber\\
-qF^{\mu\nu}p_{\nu}f^{0}\left(1+A_{0}f^{0}\right)\frac{\partial \phi_{p}}{\partial p^{\mu}}.
\eea 
The explicit expression for $\phi$ in this case is,
\bea
\phi=\sum_{n=0}^{2}\mathcal{Y}_{n}C_{\left(n\right)}^{\alpha\beta}p_{\alpha}\left(T^{-1}\partial_{\beta}T-DU_{\beta}\right).      
\eea 
Substituting it in Eq.~\eqref{eq:BUUheat}, using $\lambda^{\mu\nu}=\sum_{n=0}^{2}\lambda_{n}C_{\left(n\right)}^{\mu\beta}$ and equating the coefficients of the thermodynamic force for heat flow we have:
\bea
\label{eq:lamb0}
\lambda_{0}&=&\frac{1}{3T}\int d\Gamma_{p}~\vec{p}^{~2}f^{0} \left(1+A_{0}f^{0}\right)\mathcal{Y}_{0}\left(p\cdot U-h\right),\\
\label{eq:lamb1}
\lambda_{1}&=&\frac{1}{3T}\int d\Gamma_{p}~\vec{p}^{~2}f^{0} \left(1+A_{0}f^{0}\right)\mathcal{Y}_{1}\left(p\cdot U-h\right),\\
\label{eq:lamb2}
\lambda_{2}&=&\frac{1}{3T}\int d\Gamma_{p}~\vec{p}^{~2}f^{0} \left(1+A_{0}f^{0}\right)\mathcal{Y}_{2}\left(p\cdot U-h\right).
\eea 
Now using the expansion $ \mathcal{Y}_{n}\left(B,\tau'\right)=\sum_{m=0}^{\infty}b_{m}^{\left(n\right)}\left(B\right)L_{m}^{\frac{3}{2}}\left(\tau'\right) $
 we find the scalar unknowns $\lambda_{(n)}$ using the properties of the projection operators (given in App.~(\ref{projection})) as:
\bea
\lambda_{0}&=&-\frac{\rho T}{m}\sum_{n=0}^{\infty}b_{n}^{\left(0\right)}\beta^{\left(0\right)}_{n} , \label{lambda01}   \\
\lambda_{1}&=&-\frac{\rho T}{2m}\sum_{n=0}^{\infty}\left(b^{\left(1\right)}_{n}\beta^{\left(1\right)}_{n}+b^{\left(2\right)}_{n}\beta^{\left(2\right)}_{n}\right),    \label{lambda11}\\
\lambda_{2}&=&-\frac{\rho T}{2m}\sum_{n=0}^{\infty}\left(b^{\left(2\right)}_{n}\beta^{\left(1\right)}_{n}-b^{\left(1\right)}_{n}\beta^{\left(2\right)}_{n}\right),         \label{lambda21}
\eea
 where $\beta^{\left(0\right)}_{n}=-\frac{\beta_{n}}{3}$, $\beta^{\left(1\right)}_{n}=-\frac{2 \beta_{n}}{3}$ and $\beta^{\left(2\right)}_{n}=0$, and the expression for $\beta_n $($n=1,2)$ is given in App.~(\ref{Appendix_Thermal}) Eq.~\eqref{eq:beta_n}. The unknown coefficients $b_{m}^{(n)}$ in the expression for $\mathcal{Y}_{n}$ are determined using Eq.\eqref{eq:BUUheat}, which leads to the following set of constraints:
\bea
\label{eq:beta0lam}
 \frac{1}{\rho}\beta^{\left(0\right)}_{n}&=&-\sum_{m=1}^{\infty}b^{\left(0\right)}_{m}b^{\left(0\right)}_{mn},\\
 \label{eq:beta1lam}
 \frac{1}{\rho}\beta^{\left(1\right)}_{n}&= & -\sum_{m=1}^{\infty}  \left[b^{\left(1\right)}_{m}\left(b^{\left(1\right)}_{mn}\right)+b^{\left(2\right)}_{m}\left(\theta^{\left(1\right)}_{mn}\right)\right],    \\
 \label{eq:beta2lam}
 0&=& - \sum_{m=1}^{\infty}  \left[b^{\left(1\right)}_{m}\left(\theta^{\left(1\right)}_{mn}\right)-b^{\left(2\right)}_{m}\left(b^{\left(1\right)}_{mn}\right)\right] .
\eea 
Here we have taken into consideration the fact that $b^{\left(2\right)}_{mn}=\theta^{\left(2\right)}_{mn}=0$ due to the fact that $C_{\left(2\right)\mu\beta}\triangle^{\mu\beta}=0$.The series in the above equation starts with $m=1$, due to the conservation of energy and momentum during collisions, for example, if either $m=0$ or $n=0$, the collision bracket in Eq.~(\ref{eqn:lambda_colli}) disappears. Truncating the expansion of $\phi_{p}$ up to desired order we calculate $\lambda_{n}$ order-by-order. The detailed calculations, and the first and second-order expressions for the thermal conductivity have been provided in App.~(\ref{Appendix_Thermal}).

\section{Numerical Results}
\label{sec:numeric}
\subsection{Temperature and field dependence of the shear viscosity}

\begin{figure}[h]
	\centering 
	\includegraphics[width=.49\textwidth,origin=0,clip]{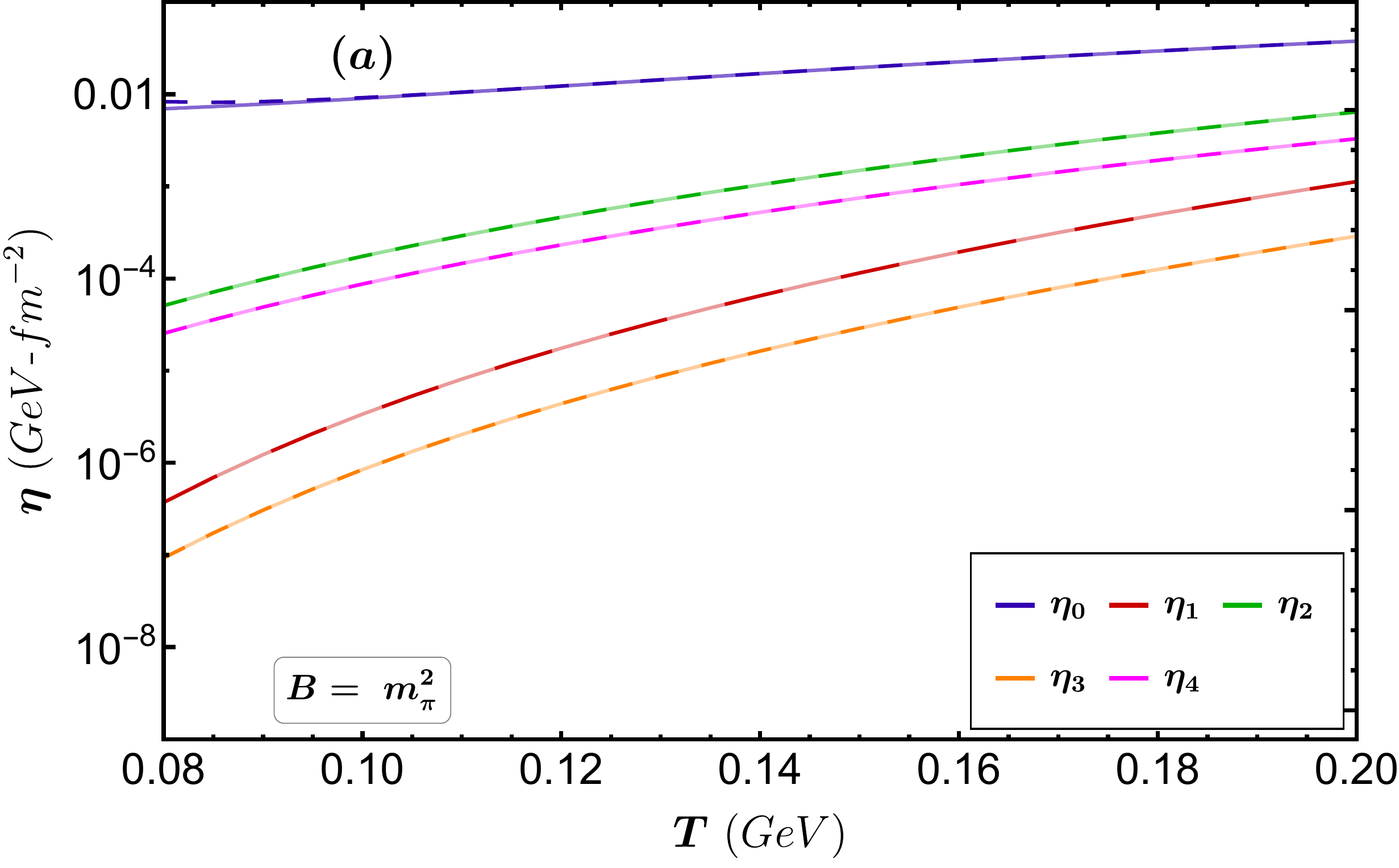}
	\hfill
	\includegraphics[width=.49\textwidth,origin=c,angle=0]{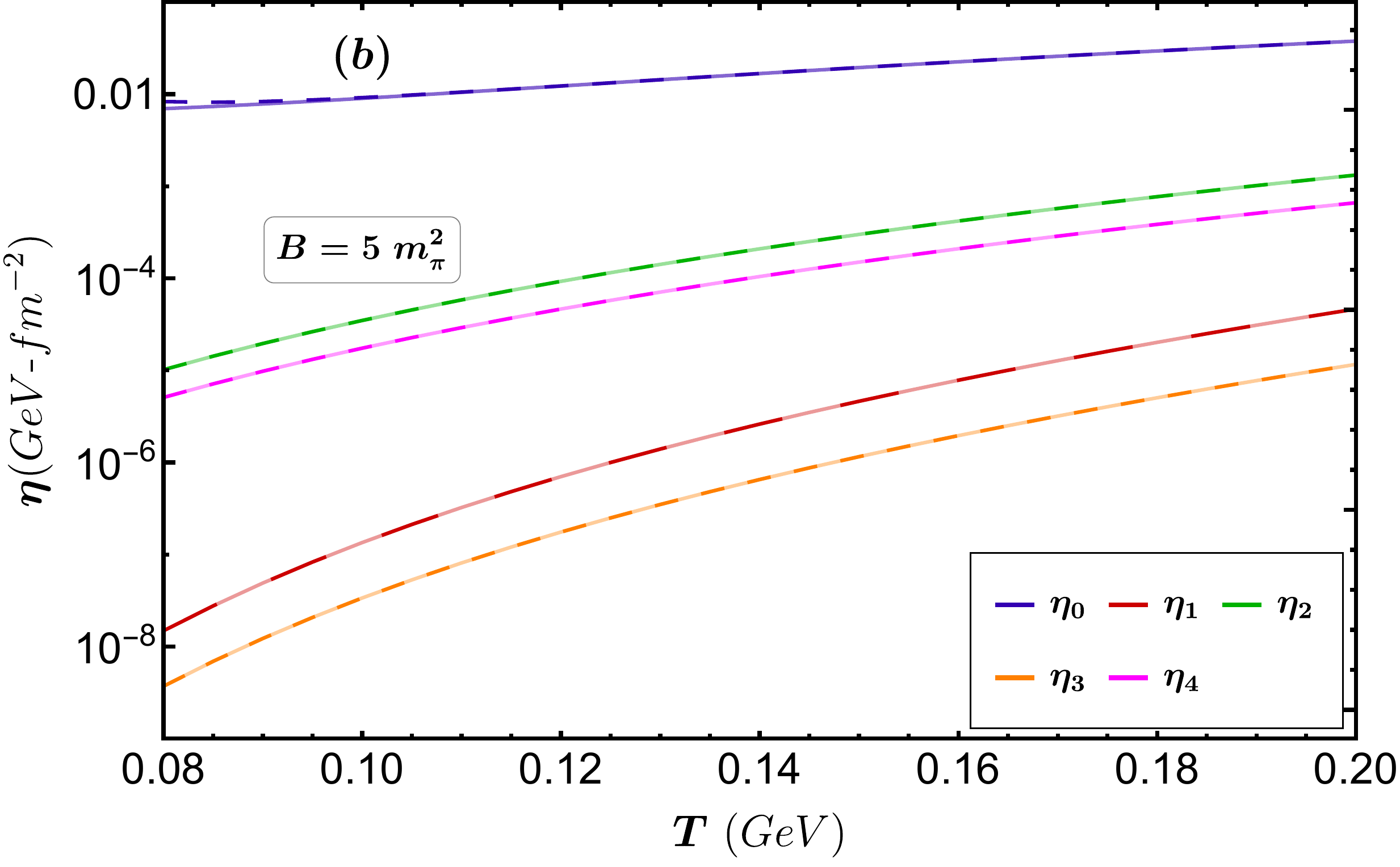}
	
	\includegraphics[width=.49\textwidth,origin=0,clip]{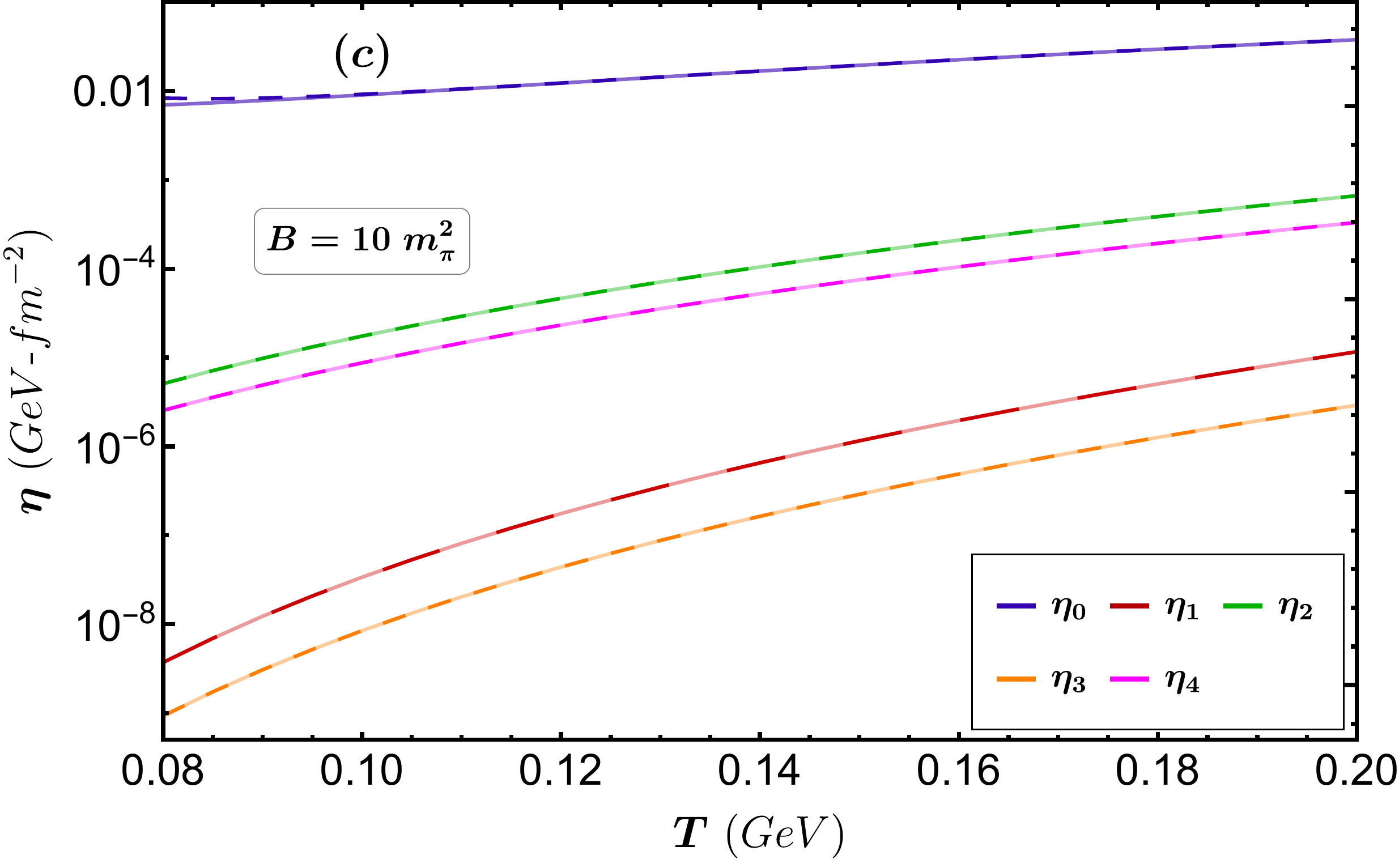}
	\hfill
	\includegraphics[width=.49\textwidth,origin=c,angle=0]{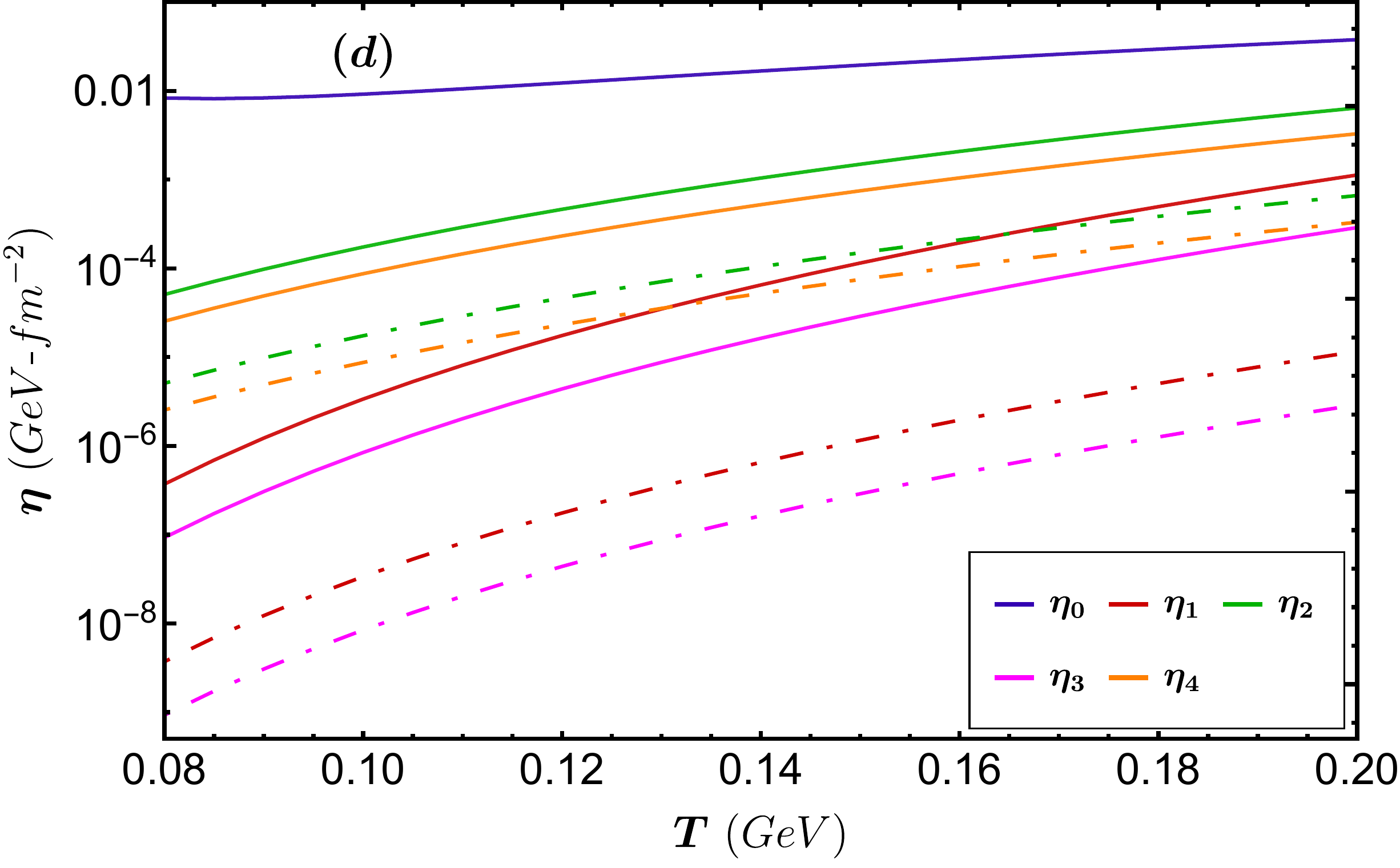}
	\caption{\label{fig:1} (Color online) $\eta_{n}$ ($\text{GeV}\text{-}fm^{-2}$) as a function of temperature, T (GeV). Panel (a), (b) and (c) corresponds to the results for magnetic field strength $m_{\pi}^{2}$, $5m_{\pi}^{2}$ and $10m_{\pi}^{2}$ respectively. The bold lines represents the first-order results ($[\eta_{n}]_{1}$), while the dashed lines correspond to the second-order results ($[\eta_{n}]_{2}$). In panel (d) we compare $[\eta_{n}]_{2}$ for $B=m_{\pi}^{2}$ (solid lines)and $B=10m_{\pi}^{2}$(dot-dashed lines). } 
\end{figure}

We intend to study the dependency of anisotropic parameters on the magnetic field order-by-order. First, we show 
the result for the shear viscosity in Fig.\eqref{fig:1}. Panel (a),(b), (c) of Fig.\eqref{fig:1} shows the temperature dependence of $\eta_{(n)}$ ($n=0\text{-}4$)
for $B=m_{\pi}^2, 5m_{\pi}^2$, and $10m_{\pi}^2$ respectively. The bold lines correspond to the first-order results, whereas the dashed lines 
represent the second-order results. The following conclusions are made: (i) all five shear viscous coefficients monotonically increase 
with temperature irrespective of the magnetic fields, (ii) $\eta_{0}$ (the parallel component) has the largest value in the temperature range considered 
here; the difference between the first-order and second-order results only visible for $\eta_{0}$ at low temperature ($\approx  80$ MeV), 
we do not see any visible difference between the first and second-order results for other $\eta$'s irrespective of the magnetic field. In panel (d) of Fig.\eqref{fig:1} we compare $\eta_{n}$ as a function of temperature for $B=m_{\pi}^{2}$ (solid lines) and $B=10 m_{\pi}^{2}$ (dot-dashed lines);
larger the magnetic fields the more anisotropic transport coefficients reduces, whereas $\eta_{0}$ is unaffected (as expected) for the all temperature ranges considered here.

\subsection{Temperature and field dependence of the bulk viscosity}

We find that the bulk viscosity does not explicitly depend on the magnetic fields, and the order-by-order results are similar to Ref.~\cite{Davesne} ; 
hence we do not discuss it in detail. However, it is interesting to note that the magnitude of $\zeta$
differs significantly order-by-order at low temperatures ($\sim$ 100 MeV) compared to the shear viscosity.

\subsection{Temperature and field dependence of the thermal conductivity}
\begin{figure}[h]
	\centering 
	\includegraphics[width=.49\textwidth,origin=0,clip]{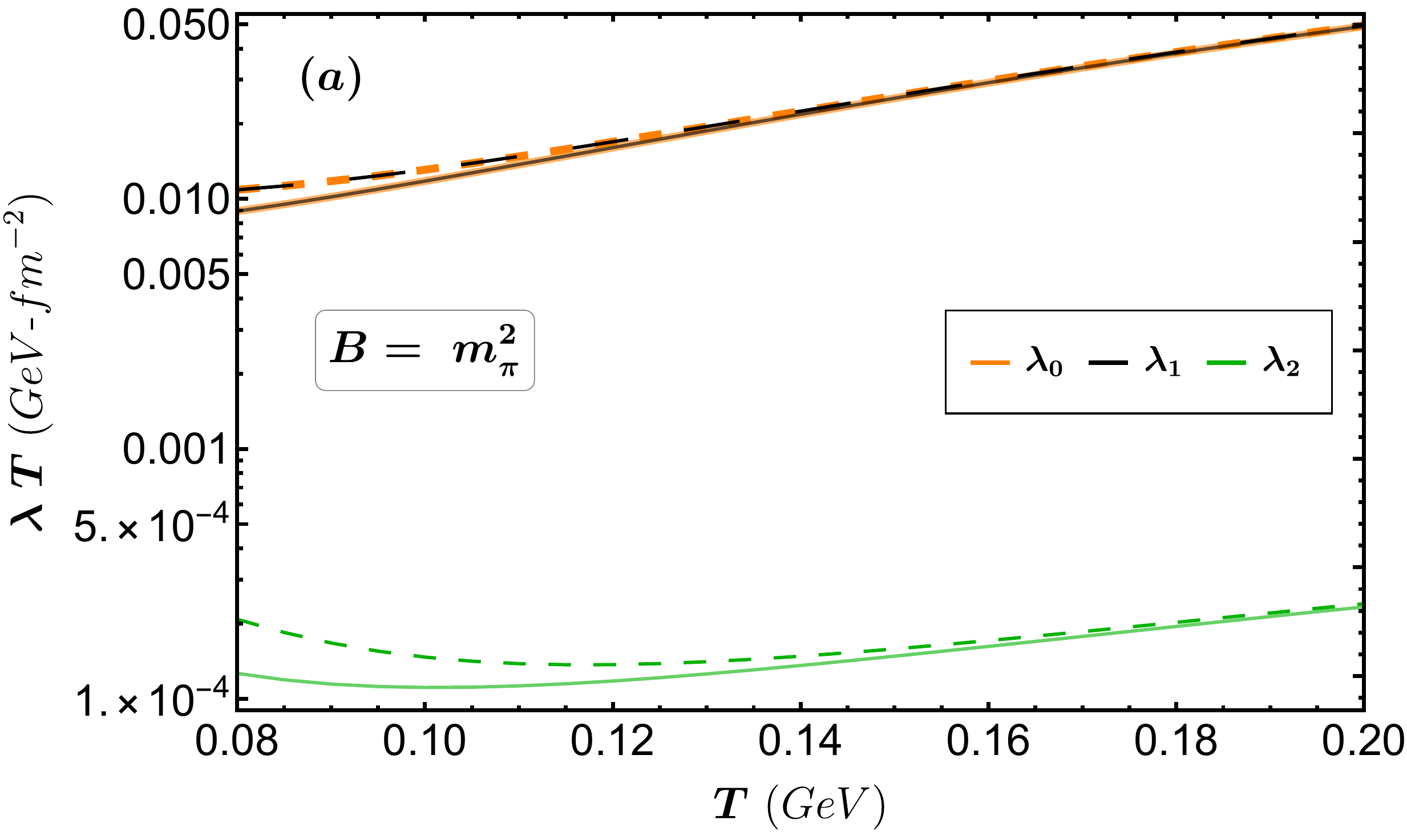}
	\hfill
	\includegraphics[width=.49\textwidth,origin=c,angle=0]{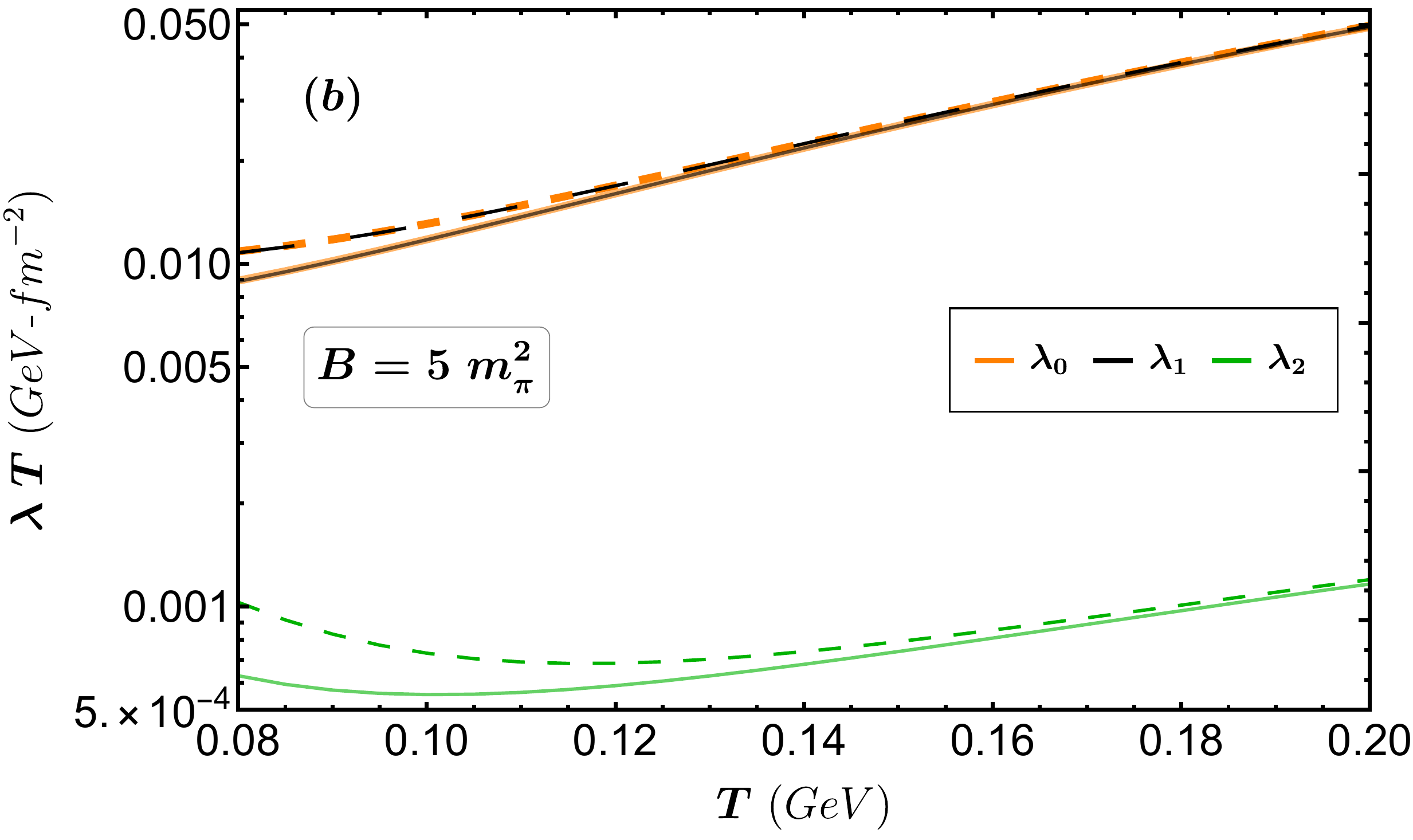}
	
	\includegraphics[width=.49\textwidth,origin=0,clip]{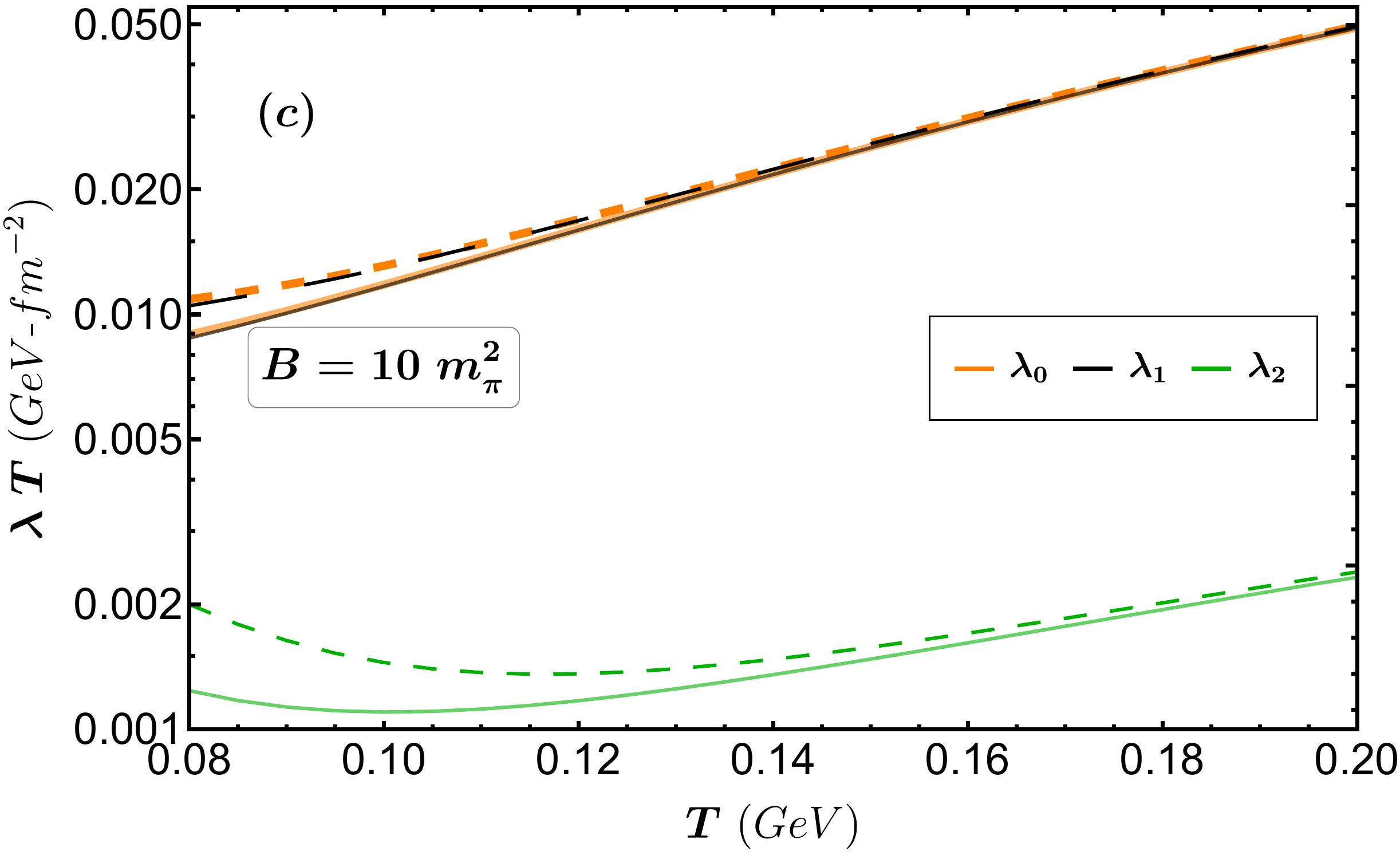}
	\hfill

	\caption{\label{fig:lambda} (Color online) $\lambda_{n}T$ ($\text{GeV}\text{-}fm^{-2}$) as a function of temperature, T (GeV). Panel (a), (b) and (c) corresponds to the results for magnetic field strength $m_{\pi}^{2}$, $5m_{\pi}^{2}$ and $10m_{\pi}^{2}$ respectively. The solid lines represent the first-order results ($[\lambda_{n}]_{1}$), while the dashed lines correspond to the second-order results ($[\lambda_{n}]_{2}$). } 
\end{figure}

As was discussed earlier, as per our tensorial decomposition of thermal conductivity tensor $\lambda^{\mu\nu}$, we have three scalar coefficients 
$\lambda_{0,1,2}$. In panel (a),(b),(c) Fig.~\eqref{fig:lambda} we show the temperature dependence of $\lambda_{n}$'s ($n=0\text{-}2$) for $B=m_{\pi}^{2}$ , $B=5m_{\pi}^{2}$, and $B=10m_{\pi}^{2}$ respectively. The solid lines in all three panels correspond to the first-order, and the dashed lines correspond to 
the second-order results. As can be seen from Fig.~\eqref{fig:lambda} $\lambda_{0}$ (orange lines) is independent of magnetic fields (it is the same as the heat coefficient for zero magnetic fields); whereas $\lambda_{1,2}$ (black and green lines) increases with magnetic fields. It is also interesting to note that 
$\left[\lambda_2\right]_1$, and $\left[\lambda_2\right]_2$ significantly differs at low temperatures ($\sim$ 100 MeV) and the difference increases with larger magnetic fields.

\section{Summary and Conclusion}
\label{sec:summ}
 In this work, we explored the magnetic field dependence of the anisotropic transport coefficients shear, bulk viscosity, and thermal conductivity 
order-by-order (up to second-order) using the relativistic Boltzmann-Uehling-Uhlenbeck transport equation. The off-equilibrium distribution function is obtained from the Chapman-Enskog method by expanding in terms of the associate Laguerre polynomial with magnetic field-dependent coefficients. 
For the current work, we only consider pions (bosons) while calculating the transport coefficients in the temperature range of 80-200 MeV.
The magnetic field varies from zero to 10$m_{\pi}^2$. Five shear viscous coefficients (some of which are non-dissipative) monotonically increase 
as a function of temperature; a visible difference between first and second-order results was seen for $\eta_{0}$ (also known as the longitudinal viscosity) at low temperature $\sim 80\text{-}100$ MeV. Other shear viscosity coefficients $\eta_{1\text{-}4}$ show no visible difference between first and second-order cases in the temperature range considered here; however, these coefficients are sensitive to the magnitude of magnetic fields. On the other hand, bulk viscous coefficients show no explicit dependence on the external magnetic field; however, they show greater sensitivity for first and second-order cases
compared to the shear viscosity in the low-temperature regime. The thermal conductivity turned out to be sensitive to both the magnetic field and the expansion order. The difference between the first- and second-order results is prominent in the low-temperature regime and is also sensitive to the external magnetic fields.

At this point, we would like to discuss a few relevant facts. Firstly, our approach must work in the low field limit as we consider the local equilibrium distribution function is the same as the zero-field case and the magnetic field enters only in the correction $\delta f$. Secondly, we assume that the electric field vanishes, or in other words, our result is applicable for large magnetic Reynolds numbers only (a.k.a. the ideal-MHD limit). Moreover, the expansion of the 
collision integral in terms of the Laguerre polynomial is not the unique choice; other choices are possible (see Ref.~\cite{Dobado:2011qu,Torres-Rincon:2012sda,Dobado:2009ek}). While calculating the collision integral, we consider the pion cross-section for vacuum; hence one can improve upon this point by taking the pionic cross-section in the medium. Last but not least similar calculations can be carried out in the future for strong-field limits.

\appendix

\section{Projection Tensors}\label{projection}

 In this appendix we give details about the projection tensor used to construct the Second $C^{\mu\nu}_{(n)}$ and fourth-rank tensor $C^{\mu\nu\alpha\beta}_{(n)}$.
$C^{\mu\nu}_{(n)}$'s are constructed from the second-rank basis projection tenor $P^{(m)}_{\mu\nu}$ which are defined as, 

\bea
P^{0}_{\mu\nu}&=&b_{\mu}b_{\nu}, \nonumber\\
P^{1}_{\mu\nu}&=&\frac{1}{2}\left(-\triangle^{\mu\nu}-b_{\mu}b_{\nu}+ib_{\mu\nu}\right),\nonumber\\
P^{-1}_{\mu\nu}&=&\frac{1}{2}\left(-\triangle^{\mu\nu}-b_{\mu}b_{\nu}-ib_{\mu\nu}\right).\nonumber
\eea 
$P^{(m)}_{\mu\nu}$ have the following properties, 
\bea
P^{(m)}_{\mu\kappa}P^{(m')\kappa}_{~~~~~~~\mu'}=\delta_{mm'}P^{(m)}_{\mu\mu'},\nonumber\\
\left(P^{(m)}_{\mu\nu}\right)^{\dagger}=P^{(-m)}_{\mu\nu}=P^{(m)}_{\nu\mu},\nonumber\\
\sum_{m=-1}^{1}P^{(m)}_{\mu\nu}=\delta_{\mu\nu};~~~P^{(m)}_{\mu\mu}=1.
\eea 
We define $C^{\mu\nu}_{(n)}$'s in terms of these basis as:
\bea
\label{eq:C0app}
C_{\left(0\right)}^{\eta\delta}&=&P^{(0)\eta\delta}, \\
\label{eq:C1app}
C_{\left(1\right)}^{\eta\delta}&=&P^{(1)\eta\delta}+P^{(-1)\eta\delta}, \\
\label{eq:C2app}
C_{\left(2\right)}^{\eta\delta}&=&i\left(P^{(1)\eta\delta}-P^{(-1)\eta\delta}\right).
\eea 

The basis for the fourth-rank tensor $C^{\mu\nu\alpha\beta}_{(n)}$'s are constructed from the  $P^{(m)}_{\mu\nu}$'s as
Using the second rank projection tensors we define the fourth rank projection tensors as,
\bea 
P^{(m)}_{\mu\nu\mu'\nu'}&=&\sum_{m_{1}=-1}^{1}\sum_{m_{2}=-1}^{1}P^{(m_{1})}_{\mu\mu'}P^{(m_{2})}_{\nu\nu'}\delta\left(m,m_{1}+m_{2}\right), \\
P^{(m)}_{\langle \mu\nu\rangle \alpha\beta}&=&P^{(m)}_{\mu\nu\alpha\beta}+P^{(m)}_{\nu\mu\alpha\beta}. \nonumber
\eea 
An important property of $P^{(m)}_{\mu\nu\mu'\nu'}$ is,
\bea
P^{(m)}_{\mu\nu\alpha\beta}P^{(n)\alpha\beta}_{~~~~~~~\mu'\nu'}=\delta\left(m,n\right)P^{\left(m\right)}_{\mu\mu'\nu\nu'}.
\eea 
Using these basis fourth-rank projection tensor, we construct $C^{\mu\nu\alpha\beta}_{(n)}$'s as:
\bea 
C_{(0)\mu\nu\alpha\beta}=P^{(0)}_{\langle \mu\nu\rangle \alpha\beta};~~~~~~~~~~~~~~~~~~~~~~~~~~~~~~~~~\nonumber\\
C_{(1)\mu\nu\alpha\beta}=P^{(1)}_{\langle \mu\nu\rangle \alpha\beta}+P^{(-1)}_{\langle \mu\nu\rangle \alpha\beta};~~~~~C_{(2)\mu\nu\alpha\beta}=i\left(P^{(1)}_{\langle \mu\nu\rangle \alpha\beta}+P^{(-1)}_{\langle \mu\nu\rangle \alpha\beta}\right);\nonumber\\
C_{(3)\mu\nu\alpha\beta}=P^{(2)}_{\langle \mu\nu\rangle \alpha\beta}+P^{(-2)}_{\langle \mu\nu\rangle \alpha\beta};~~~~~C_{(4)\mu\nu\alpha\beta}=i\left(P^{(2)}_{\langle \mu\nu\rangle \alpha\beta}+P^{(-2)}_{\langle \mu\nu\rangle \alpha\beta}\right).\nonumber
\eea 
One can further verify that 
\begin{equation}
C_{(0)\mu\nu\alpha\beta}+C_{(1)\mu\nu\alpha\beta}+C_{(3)\mu\nu\alpha\beta}=\triangle_{\mu\alpha}\triangle_{\nu\beta}+\triangle_{\nu\alpha}\triangle_{\mu\beta}.
\end{equation}

\section{ Detail Derivation of the Shear Viscosity}\label{Appendix_Shear}

In this appendix we give the details of the results presented in Sec.~(\ref{sec:shear}).
We start with the right hand side of the BUU equation \eqref{eq:ShearBUU} and denote the two terms as $T_{1}$, and $T_{2}$, where
\bea
T_{1}=\int d\Gamma_{k}d\Gamma_{p'}d\Gamma_{k'}\big[f^{0}(x,p)f^{0}(x,k)\left(1+A_{0}f^{0}(x,p')\right)\left(1+A_{0}f^{0}(x,k')\right)~~~~~\nonumber\\
\times\left(\phi(x,p)+\phi(x,k)-\phi(x,p')
-\phi(x,k')\right)W\big]
\eea
and,
\bea
T_{2}&=&qBb^{\mu\nu}p_{\nu}f^{0}\left(1+A_{0}f^{0}\right)\frac{\partial \phi_{p}}{\partial p^{\mu}} \nonumber\\
&=&qBf^{0}\left(1+A_{0}f^{0}\right)b^{\mu\nu}p_{\nu}\frac{\partial}{\partial p^{\mu}}\left[\sum_{n=1}^{4}\mathcal{X}_{n}C_{\left(
	n\right)\alpha\beta\eta\delta}\langle p^{\alpha}p^{\beta}\rangle V^{\eta\delta}\right]\nonumber\\
&=&qBf^{0}\left(1+A_{0}f^{0}\right)b^{\mu\nu}p_{\nu}\Delta^{\alpha\beta}_{~~~\kappa\delta}\left[\sum_{n=1}^{4}C_{\left(
	n\right)\alpha\beta\eta\delta}p^{\kappa}p^{\delta}\frac{\partial \mathcal{X}_{n}}{\partial p^{\mu}}+\sum_{n=1}^{4}\mathcal{X}_{n}C_{\left(
	n\right)\alpha\beta\eta\delta}\frac{\partial \left( p^{\kappa}p^{\delta}\right)}{\partial p^{\mu}}\right]V^{\eta\delta}. \nonumber
\eea 
The first term of the above equation vanishes because,
\bea
\frac{\partial \mathcal{X}_{n}}{\partial p^{\mu}}=\mathcal{X}'_{n}\frac{\partial }{\partial p^{\mu}}\left(\frac{p\cdot U}{T}\right)=\frac{\mathcal{X}'_{n}}{T}U_{\mu};~~~~ b^{\mu\nu}U_{\nu}=0,
\eea 
where we have used Eq.~\eqref{eq:Xn} for $\mathcal{X}_{n}$.  Thus,
\bea
T_{2}=-qBf^{0}\left(1+A_{0}f^{0}\right)\Delta^{\alpha\beta}_{~~~\kappa\delta}p^{\nu}\left(b^{\nu\kappa}p^{\delta}+b^{\nu\delta}p^{\kappa}\right) \left[\sum_{n=1}^{4}\mathcal{X}_{n}C_{\left(n\right)\alpha\beta\eta\delta}\right]V^{\eta\delta}.
\eea 
We further simply the above equation and express in terms of $\langle p^{\mu}p^{\nu}\rangle$  using the relation $-\frac{1}{2}\left(b^{\mu\kappa}\triangle^{\nu\delta}+b^{\mu\delta}\triangle^{\nu\kappa}+b^{\nu\kappa}\triangle^{\mu\delta}+b^{\nu\delta}\triangle^{\mu\kappa}\right)=\frac{1}{2}C_{\left(2\right)}^{\mu\nu\kappa\delta}+C_{\left(4\right)}^{\mu\nu\kappa\delta}$, and using the properties of the basis projection tensors, 
\bea
T_{2}=2qBf^{0}\left(1+f^{0}\right)V_{\eta\delta}\langle p_{\mu}p_{\nu}\rangle\Big[-\frac{\mathcal{X}_{2}}{2}C_{\left(1\right)}^{\mu\nu\eta\delta}+\frac{\mathcal{X}_{1}}{2}C_{\left(2\right)}^{\mu\nu\eta\delta}-\mathcal{X}_{4}C_{\left(3\right)}^{\mu\nu\eta\delta}+\mathcal{X}_{3}C_{\left(4\right)}^{\mu\nu\eta\delta}\Big].
\eea 
Our goal is to find the unknown coefficients $\mathcal{X}_{n}$, for that purpose we use the above form of the collision integral in the BUU equation Eq.\eqref{eq:ShearBUU}  and contract both sides with $P^{\eta\delta\alpha\beta}_{\left(n\right)}$; after that we integrate with a weight function $L^{\frac{5}{2}}_{j}\left(\tau_{p}\right)\langle p_{\alpha}p_{\beta}\rangle$ and get Eqs.~(\ref{c0_relation})-(\ref{c2_relation}). Various terms in these equations are defined as, 
\bea
\gamma^{\left(n\right)}_{j}&=&\frac{1}{\rho T^{2}}\int d\Gamma_{p}f^{0}\left(1+A_{0}f^{0}\right)\langle p_{\alpha}p_{\beta}\rangle \langle p_{\eta}p_{\delta}\rangle L^{\frac{5}{2}}_{j}\left(\tau_{p}\right)C_{\left(n\right)}^{\alpha\beta\eta\delta}, \nonumber\\    
&=&\frac{1}{10}\Delta_{\alpha\beta\psi\theta}\Delta_{\eta\delta\kappa\pi}C_{\left(n\right)}^{\alpha\beta\eta\delta}\left[\Delta^{\psi\theta}\Delta^{\kappa\pi}+\Delta^{\psi\kappa}\Delta^{\theta\pi}+\Delta^{\psi\pi}\Delta^{\theta\kappa}\right]\gamma_{j},  \label{eq:gammajn}    \\
\label{eq:appcnj}
c_{nj}&=&\frac{1}{m^{2}T^{2}} \left[L^{\frac{5}{2}}_{n}\langle p_{\mu}p_{\nu}\rangle ,L^{\frac{5}{2}}_{j}\langle p_{\eta}p_{\delta}\rangle \right] _{C_{\left(0\right)}},    \nonumber\\
&=&\frac{1}{10}\Delta_{\alpha\beta\psi\theta}\Delta_{\eta\delta\kappa\pi}C_{\left(0\right)}^{\alpha\beta\eta\delta}\left[\Delta^{\psi\theta}\Delta^{\kappa\pi}+\Delta^{\psi\kappa}\Delta^{\theta\pi}+\Delta^{\psi\pi}\Delta^{\theta\kappa}\right]\gamma_{j}c'_{nj}, \\
\label{eq:appdnj}
d_{nj}&=&\frac{1}{m^{2}T^{2}} \left[L^{\frac{5}{2}}_{n}\langle p_{\mu}p_{\nu}\rangle ,L^{\frac{5}{2}}_{j}\langle p_{\eta}p_{\delta}\rangle \right] _{C_{\left(1\right)}}, \nonumber  \\
&=&\frac{1}{10}\Delta_{\alpha\beta\psi\theta}\Delta_{\eta\delta\kappa\pi}C_{\left(1\right)}^{\alpha\beta\eta\delta}\left[\Delta^{\psi\theta}\Delta^{\kappa\pi}+\Delta^{\psi\kappa}\Delta^{\theta\pi}+\Delta^{\psi\pi}\Delta^{\theta\kappa}\right]\gamma_{j}c'_{nj}, \\
\label{eq:applnj}
l_{nj}&=&\frac{1}{m^{2}T^{2}} \left[L^{\frac{5}{2}}_{n}\langle p_{\mu}p_{\nu}\rangle ,L^{\frac{5}{2}}_{j}\langle p_{\eta}p_{\delta}\rangle \right] _{C_{\left(3\right)}},    \nonumber\\
&=&\frac{1}{10}\Delta_{\alpha\beta\psi\theta}\Delta_{\eta\delta\kappa\pi}C_{\left(3\right)}^{\alpha\beta\eta\delta}\left[\Delta^{\psi\theta}\Delta^{\kappa\pi}+\Delta^{\psi\kappa}\Delta^{\theta\pi}+\Delta^{\psi\pi}\Delta^{\theta\kappa}\right]\gamma_{j}c'_{nj}, \\
\label{eq:xi}
\xi_{nj}^{\left(a\right)}&=&\frac{qB}{\rho^{2}T^{2}}\int d\Gamma_{p}f^{0}\left(1+A_{0}f^{0}\right)\langle p_{\alpha}p_{\beta}\rangle \langle p_{\eta}p_{\delta}\rangle C_{\left(a\right)}^{\alpha\beta\eta\delta}L_{n}^{\frac{5}{2}}\left(\tau'\right)L_{j}^{\frac{5}{2}}\left(\tau'\right).
\eea

Where $\gamma_{n}$ with only subscript corresponds to the one mentioned in Ref.~\cite{Davesne}, $c'_{nj}$ corresponds to the quantity $c_{nj}$ in the above mentioned paper, and
\bea
\left[\mathcal{K}_{m}\langle p_{\mu}p_{\nu}\rangle ,\mathcal{F}_{j}\langle p_{\eta}p_{\delta}\rangle \right]_{C_{\left(l\right)}}=\frac{m^{2}}{4\rho^{2}}\int d\Gamma_{p}d\Gamma_{k}d\Gamma_{p'}d\Gamma_{k'}\Big[f^{0}_{p}f^{0}_{k}\left(1+A_{0}f^{0}_{p'}\right)\left(1+A_{0}f^{0}_{k'}\right)\nonumber\\
\times \left(\mathcal{K}_{m}\left(\tau_{p}\right)\langle p_{\mu}p_{\nu}\rangle +\mathcal{K}_{m}\left(\tau_{k}\right)\langle k_{\mu}k_{\nu}\rangle -\mathcal{K}_{m}\left(\tau_{p'}\right)\langle p'_{\mu}p'_{\nu}\rangle -\mathcal{K}_{m}\left(\tau_{k'}\right)\langle k'_{\mu}k'_{\nu}\rangle \right) \nonumber\\
\times \left(\mathcal{F}_{j}\left(\tau_{p}\right)\langle p_{\mu}p_{\nu}\rangle +\mathcal{F}_{j}\left(\tau_{k}\right)\langle k_{\mu}k_{\nu}\rangle -\mathcal{F}_{j}\left(\tau_{p'}\right)\langle p'_{\mu}p'_{\nu}\rangle -\mathcal{F}_{j}\left(\tau_{k'}\right)\langle k'_{\mu}k'_{\nu}\rangle \right)C_{\left(l\right)}^{\mu\nu\eta\delta}W\Big].   ~~~   
\eea 
In Eq.~(\ref{c1_relation}) and Eq.~(\ref{c2_relation}) we also get terms like Eqs.~\eqref{eq:appcnj},\eqref{eq:appdnj},\eqref{eq:applnj} for $C_{(2)}$, $C_{(4)}$, and 
$\xi_{nj}^{\left(2\right)}$ and $\xi_{nj}^{\left(4\right)}$ but on careful observation one see that these terms vanishes because   $C_{\left(2\right)}^{\alpha\beta\eta\delta}\left[\Delta^{\alpha\beta}\Delta^{\eta\delta}+\Delta^{\alpha\eta}\Delta^{\beta\delta}+\Delta^{\alpha\delta}\Delta^{\beta\eta}\right]=C_{\left(4\right)}^{\alpha\beta\eta\delta}\left[\Delta^{\alpha\beta}\Delta^{\eta\delta}+\Delta^{\alpha\eta}\Delta^{\beta\delta}+\Delta^{\alpha\delta}\Delta^{\beta\eta}\right]=0$.  Once we evaluate the collision integral as described above we further proceed to calculate the shear viscous coefficients 
from the definition:
\begin{equation}
\pi^{\mu\nu}=\eta^{\mu\nu\psi\theta}\langle\partial_{\psi}U_{\theta} \rangle
=\int d\Gamma \langle p^{\mu}p^{\nu}\rangle f^{0}\left(1+A_{0}f^{0}\right)\sum_{n=0}^{4}\mathcal{X}_{n}C_{\left(n\right)\alpha\beta\eta\delta}p^{\alpha}p^{\beta}V^{\eta\delta}
\end{equation}
Now using the definition Eq.~\eqref{eq:shearfourth} in the above expression and equating the coefficient of $\partial_{\lambda}U_{\kappa}$ we have:
\bea
\sum_{n=0}^{4}\eta_{n}C_{\left(n\right)}^{\mu\nu\psi\theta}\left[\triangle^{\lambda}_{\psi}\triangle^{\kappa}_{\theta}+\triangle^{\kappa}_{\psi}\triangle^{\lambda}_{\theta}-\frac{2}{3}\triangle^{\lambda\kappa}\triangle_{\psi\theta}\right]= ~~~~~~~~~~~~~~~~~~~~~~~~~~~~~~~~~~~  ~~~~~     \nonumber\\
\int d\Gamma \langle p^{\mu}p^{\nu}\rangle f^{0}\left(1+A_{0}f^{0}\right)\sum_{n=0}^{4}\mathcal{X}_{n}C_{\left(n\right)\alpha\beta\eta\delta}p^{\alpha}p^{\beta} \left[\triangle^{\eta\lambda}\triangle^{\delta\kappa}+\triangle^{\eta\kappa}\triangle^{\delta\lambda}\right]       
\eea
where $V^{\eta\delta} =\frac{1}{2}\left[\triangle^{\eta\lambda}\triangle^{\delta\kappa}+\triangle^{\eta\kappa}\triangle^{\delta\lambda}\right]\partial_{\lambda}U_{\kappa} $.  Finally we obtain the results $\left[\eta_{n}\right]$ for arbitrary orders ( Eqs. (\ref{eta0}), (\ref{eta12}), (\ref{eta34})) by contracting both sides with $P^{\left(n\right)}_{\lambda\kappa\psi\theta}$. If we keep terms up to first-order in $\mathcal{X}$, we have the first-order results given in Eq.~\eqref{eq:eta01}-\eqref{eq:eta41}. 
Similarly by keeping terms up to second in the coefficient $\mathcal{X}_{n}$ we get,
\bea
\left[\eta_{0}\right]_{2}&=&\frac{T}{4}\frac{\left[c_{11}\left(\gamma_{1}^{\left(0\right)}\right)^{2}-c_{01}\gamma_{0}^{\left(0\right)}\gamma_{1}^{\left(0\right)}-c_{10}\gamma_{0}^{\left(0\right)}\gamma_{1}^{\left(0\right)}+c_{00}\left(\gamma_{1}^{\left(0\right)}\right)^{2}\right]}{\left(c_{00}c_{11}-c_{01}c_{10}\right)}  \\
\left[\eta_{1}\right]_{2}&=&\frac{\rho T^{2}}{4}\left[c_{0}^{\left(1\right)}\gamma_{0}^{\left( 1\right)}+c_{1}^{\left(1\right)}\gamma_{1}^{\left( 1\right)}\right],   \\
\left[\eta_{2}\right]_{2}&=&\frac{\rho T^{2}}{4}\left[c_{0}^{\left(2\right)}\gamma_{0}^{\left( 1\right)}+c_{1}^{\left(2\right)}\gamma_{1}^{\left( 1\right)}\right],\\
\left[\eta_{3}\right]_{2}&=&\frac{\rho T^{2}}{4}\left[c_{0}^{\left(3\right)}\gamma_{0}^{\left( 3\right)}+c_{1}^{\left(3\right)}\gamma_{1}^{\left( 3\right)}\right],   \\
\left[\eta_{4}\right]_{2}&=&\frac{\rho T^{2}}{4}\left[c_{0}^{\left(4\right)}\gamma_{0}^{\left( 3\right)}+c_{1}^{\left(4\right)}\gamma_{1}^{\left( 3\right)}\right].
\eea
Where using Eq.~\eqref{c1_relation}, \eqref{c2_relation} we have,
\bea
c^{\left(1\right)}_{0}&=&\frac{1}{2\rho T D}\Bigg[ \gamma_{0}^{\left(1\right)} d_{00} d_{11}^{2} -\left(d_{10} \left(\gamma_{0}^{\left(1\right)} d_{01}+\gamma_{1}^{\left(1\right)} d_{00}\right)   +\xi_{10}^{\left(1\right)} \left(\gamma_{1}^{\left(1\right)} \xi_{00}^{\left(1\right)}-\gamma_{0}^{\left(1\right)} \xi_{01}^{\left(1\right)}\right)\right)d_{11}        \nonumber \\
& &-\gamma_{0}^{\left(1\right)} d_{01} \xi_{11}^{\left(1\right)} \xi_{10}^{\left(1\right)} -\gamma_{0}^{\left(1\right)} d_{10} \xi_{11}^{\left(1\right)} \xi_{01}^{\left(1\right)}   +\gamma_{1}^{\left(1\right)} d_{01} \left(\xi_{10}^{\left(1\right)}\right)^{2}  \nonumber \\
& &+\gamma_{1}^{\left(1\right)} d_{10}^{2} d_{01} +  \gamma_{1}^{\left(1\right)} d_{10} \xi_{11}^{\left(1\right)} \xi_{00}^{\left(1\right)} +  d_{00} \xi_{11}^{\left(1\right)} \left(\gamma_{0}^{\left(1\right)} \xi_{11}^{\left(1\right)}-\gamma_{1}^{\left(1\right)} \xi_{10}^{\left(1\right)}\right) \Bigg], \\
c^{\left(1\right)}_{1}&=&\frac{1}{2\rho T D}\Bigg[-\left(d_{00} \xi_{01}^{\left(1\right)}-d_{01} \xi_{00}^{\left(1\right)}\right) \left(\gamma_{0}^{\left(1\right)} \xi_{11}^{\left(1\right)}-\gamma_{1}^{\left(1\right)} \xi_{10}^{\left(1\right)}\right)  \nonumber\\
& &+d_{11} \left(-\gamma_{0}^{\left(1\right)} d_{01} d_{00}+\gamma_{1}^{\left(1\right)} \left(d_{00}^{2}+\left(\xi_{00}^{\left(1\right)}\right)^{2}\right)-\gamma_{0}^{\left(1\right)} \xi_{01}^{\left(1\right)} \xi_{00}^{\left(1\right)}\right)      \nonumber\\
& &+d_{10} \left(\gamma_{0}^{\left(1\right)} d_{01}^2-\gamma_{1}^{\left(1\right)} d_{00} d_{01}+\xi_{01} \left(\gamma_{0}^{\left(1\right)} \xi_{01}^{\left(1\right)}-\gamma_{1}^{\left(1\right)} \xi_{00}^{\left(1\right)}\right)\right) \Bigg], \\
c^{\left(2\right)}_{0}&=&\frac{1}{2\rho T D}\Big[ d_{11} \xi_{10}^{\left(1\right)} \left(\gamma_{1}^{\left(1\right)} d_{00}-\gamma_{0}^{\left(1\right)} d_{01}\right)+\gamma_{0}^{\left(1\right)} d_{11}^{2} \xi_{00}^{\left(1\right)}+\gamma_{1}^{\left(1\right)} d_{10}^{2} \xi_{01}^{\left(1\right)}     \nonumber\\
& &+\left(\xi_{10}^{\left(1\right)} \xi_{01}^{\left(1\right)}-\xi_{11}^{\left(1\right)} \xi_{00}^{\left(1\right)}\right) \left(\gamma_{1}^{\left(1\right)} \xi_{10}^{\left(1\right)}-\gamma_{0}^{\left(1\right)} \xi_{11}^{\left(1\right)}\right)       \nonumber\\
& &+d_{10} \left(-d_{11} \left(\gamma_{0}^{\left(1\right)} \xi_{01}^{\left(1\right)}+\gamma_{1}^{\left(1\right)} \xi_{00}^{\left(1\right)}\right)+\gamma_{0}^{\left(1\right)} d_{01} \xi_{11}^{\left(1\right)}-\gamma_{1}^{\left(1\right)} d_{00} \xi_{11}^{\left(1\right)}\right)  \Big], \\ 
c^{\left(2\right)}_{1}&=&\frac{1}{2\rho T D}\Big[ \gamma_{0}^{\left(1\right)} d_{01}^{2} \xi_{10}^{\left(1\right)} + \gamma_{0}^{\left(1\right)} d_{11} d_{00} \xi_{01}^{\left(1\right)}+\gamma_{0}^{\left(1\right)} \xi_{10}^{\left(1\right)} \left(\xi_{01}^{\left(1\right)}\right)^2-\gamma_{0}^{\left(1\right)} \xi_{11}^{\left(1\right)} \xi_{00}^{\left(1\right)} \xi_{01}^{\left(1\right)}   \nonumber\\
& &-d_{01} \left(d_{00} \left(\gamma_{0}^{\left(1\right)} \xi_{11}^{\left(1\right)}+\gamma_{1}^{\left(1\right)} \xi_{10}^{\left(1\right)}\right)+\xi_{00}^{\left(1\right)} \left(\gamma_{0}^{\left(1\right)} d_{11}-\gamma_{1}^{\left(1\right)} d_{10}\right)\right)     \nonumber\\
& &+\gamma_{1}^{\left(1\right)} d_{00}^{2} \xi_{11}^{\left(1\right)}-\gamma_{1}^{\left(1\right)} d_{10} d_{00} \xi_{01}^{\left(1\right)}+\gamma_{1}^{\left(1\right)} \xi_{11}^{\left(1\right)} \left(\xi_{00}^{\left(1\right)}\right)^{2}-\gamma_{1}^{\left(1\right)} \xi_{10}^{\left(1\right)} \xi_{01}^{\left(1\right)} \xi_{00}^{\left(1\right)}  \Big],
\eea
The determinant $D$ in the above expressions is:
\bea
D&=&\Big[\left(d^{2}_{00}+\left(\xi_{00}^{\left(1\right)}\right)^{2}\right)d^{2}_{11}+2\xi_{10}^{\left(1\right)}\left(d_{00}\xi_{01}^{\left(1\right)}-d_{01} \xi_{00}^{\left(1\right)}\right)d_{11}+d_{00}^{2} \left(\xi_{11}^{\left(1\right)}\right)^{2}+d_{01}^{2} \left(\xi_{10}^{\left(1\right)}\right)^{2}          \nonumber   \\
& &+\left(\xi_{10}^{\left(1\right)}\xi_{01}^{\left(1\right)}\right)^2+\left(\xi_{00}^{\left(1\right)} \xi_{11}^{\left(1\right)}\right)^{2} -2 d_{01} d_{00} \xi_{11}^{\left(1\right)} \xi_{10}^{\left(1\right)}   +d_{10}^2 \left(d_{01}^{2}+\left(\xi_{01}^{\left(1\right)}\right)^{2}\right)-2 \xi_{11}^{\left(1\right)} \xi_{10}^{\left(1\right)} \xi_{01}^{\left(1\right)} \xi_{00}^{\left(1\right)}                 \nonumber  \\
& &-2 d_{10} \left(\xi_{11}^{\left(1\right)} \left(d_{00} \xi_{01}^{\left(1\right)}-d_{01} \xi_{00}^{\left(1\right)}\right)+d_{11} \left(d_{01} d_{00}+\xi_{01}^{\left(1\right)} \xi_{00}^{\left(1\right)}\right)\right)        \Big].
\eea 
Similarly the other coefficients are given by 
\bea
c^{\left(3\right)}_{0}&=&\frac{1}{2\rho T D'}\Bigg[ \gamma_{0}^{\left(3\right)} l_{00} l_{11}^{2} -\left(l_{10} \left(\gamma_{0}^{\left(3\right)} l_{01}+\gamma_{1}^{\left(3\right)} l_{00}\right)   +4\xi_{10}^{\left(3\right)} \left(\gamma_{1}^{\left(3\right)} \xi_{00}^{\left(3\right)}-\gamma_{0}^{\left(3\right)} \xi_{01}^{\left(3\right)}\right)\right)l_{11}        \nonumber \\
& &-4\gamma_{0}^{\left(3\right)} l_{01} \xi_{11}^{\left(3\right)} \xi_{10}^{\left(3\right)}     -4\gamma_{0}^{\left(3\right)} l_{10} \xi_{11}^{\left(3\right)} \xi_{01}^{\left(3\right)}   +4\gamma_{1}^{\left(3\right)} l_{01} \left(\xi_{10}^{\left(3\right)}\right)^{2}     \nonumber \\
& &+\gamma_{1}^{\left(3\right)} l_{10}^{2} l_{01} + 4 \gamma_{1}^{\left(3\right)} l_{10} \xi_{11}^{\left(3\right)} \xi_{00}^{\left(3\right)} + 4 l_{00} \xi_{11}^{\left(3\right)} \left(\gamma_{0}^{\left(3\right)} \xi_{11}^{\left(3\right)}-\gamma_{1}^{\left(3\right)} \xi_{10}^{\left(3\right)}\right) \Bigg], \\
c^{\left(3\right)}_{1}&=&\frac{1}{2\rho T D'}\Bigg[-4\left(l_{00} \xi_{01}^{\left(3\right)}-l_{01} \xi_{00}^{\left(3\right)}\right) \left(\gamma_{0}^{\left(3\right)} \xi_{11}^{\left(3\right)}-\gamma_{1}^{\left(3\right)} \xi_{10}^{\left(3\right)}\right)  \nonumber\\
& &+l_{11} \left(-\gamma_{0}^{\left(3\right)} l_{01} l_{00}+\gamma_{1}^{\left(3\right)} \left(l_{00}^{2}+\left(2\xi_{00}^{\left(3\right)}\right)^{2}\right)-4\gamma_{0}^{\left(3\right)} \xi_{01}^{\left(3\right)} \xi_{00}^{\left(3\right)}\right)      \nonumber\\
& &+l_{10} \left(\gamma_{0}^{\left(3\right)} l_{01}^2-\gamma_{1}^{\left(3\right)} l_{00} l_{01}+4\xi_{01}^{\left(3\right)} \left(\gamma_{0}^{\left(3\right)} \xi_{01}^{\left(3\right)}-\gamma_{1}^{\left(3\right)} \xi_{00}^{\left(3\right)}\right)\right) \Bigg], \\
c^{\left(4\right)}_{0}&=&\frac{1}{2\rho T D'}\Big[ 2l_{11} \xi_{10}^{\left(3\right)} \left(\gamma_{1}^{\left(3\right)} l_{00}-\gamma_{0}^{\left(3\right)} l_{01}\right)+2\gamma_{0}^{\left(3\right)} l_{11}^{2} \xi_{00}^{\left(3\right)}+2\gamma_{1}^{\left(3\right)} l_{10}^{2} \xi_{01}^{\left(3\right)}     \nonumber\\
& & +8\left(\xi_{10}^{\left(3\right)} \xi_{01}^{\left(3\right)}-\xi_{11}^{\left(3\right)} \xi_{00}^{\left(3\right)}\right) \left(\gamma_{1}^{\left(3\right)} \xi_{10}^{\left(3\right)}-\gamma_{0}^{\left(3\right)} \xi_{11}^{\left(3\right)}\right)       \nonumber\\
& &+2l_{10} \left(-l_{11} \left(\gamma_{0}^{\left(3\right)} \xi_{01}^{\left(3\right)}+\gamma_{1}^{\left(3\right)} \xi_{00}^{\left(3\right)}\right)+\gamma_{0}^{\left(3\right)} l_{01} \xi_{11}^{\left(3\right)}-\gamma_{1}^{\left(3\right)} l_{00} \xi_{11}^{\left(3\right)}\right)  \Big], \\
c^{\left(4\right)}_{1}&=&\frac{1}{2\rho T D'}\Big[ 2 \gamma_{0}^{\left(3\right)} l_{01}^{2} \xi_{10}^{\left(3\right)} + 2 \gamma_{0}^{\left(3\right)} l_{11} l_{00} \xi_{01}^{\left(3\right)}+8\gamma_{0}^{\left(3\right)} \xi_{10}^{\left(3\right)} \left(\xi_{01}^{\left(3\right)}\right)^2-8\gamma_{0}^{\left(3\right)} \xi_{11}^{\left(3\right)} \xi_{00}^{\left(3\right)} \xi_{01}^{\left(3\right)}   \nonumber\\
& & -2l_{01} \left(l_{00} \left(\gamma_{0}^{\left(3\right)} \xi_{11}^{\left(3\right)}+\gamma_{1}^{\left(3\right)} \xi_{10}^{\left(3\right)}\right)+\xi_{00}^{\left(3\right)} \left(\gamma_{0}^{\left(3\right)} l_{11}-\gamma_{1}^{\left(3\right)} l_{10}\right)\right)     \nonumber\\
& & +2\gamma_{1}^{\left(3\right)} l_{00}^{2} \xi_{11}^{\left(3\right)}-2\gamma_{1}^{\left(3\right)} l_{10} l_{00} \xi_{01}^{\left(3\right)}+8\gamma_{1}^{\left(3\right)} \xi_{11}^{\left(3\right)} \left(\xi_{00}^{\left(3\right)}\right)^{2}-8\gamma_{1}^{\left(3\right)} \xi_{10}^{\left(3\right)} \xi_{01}^{\left(3\right)} \xi_{00}^{\left(3\right)}  \Big],
\eea
The determinant $D'$ in the above expressions is 
\bea
D'&=&\Big[\left(l^{2}_{00}+\left(2\xi_{00}^{\left(3\right)}\right)^{2}\right)l^{2}_{11}+8\xi_{10}^{\left(3\right)}\left(l_{00}\xi_{01}^{\left(3\right)}-l_{01} \xi_{00}^{\left(3\right)}\right)l_{11}+4l_{00}^{2} \left(\xi_{11}^{\left(3\right)}\right)^{2}+4l_{01}^{2} \left(\xi_{10}^{\left(3\right)}\right)^{2}          \nonumber   \\
& & +16\left(\xi_{10}^{\left(3\right)}\xi_{01}^{\left(3\right)}\right)^2+16\left(\xi_{00}^{\left(3\right)} \xi_{11}^{\left(3\right)}\right)^{2} -8 l_{01} l_{00} \xi_{11}^{\left(3\right)} \xi_{10}^{\left(3\right)}   +l_{10}^2 \left(l_{01}^{2}+4\left(\xi_{01}^{\left(3\right)}\right)^{2}\right) \nonumber  \\
& & -32 \xi_{11}^{\left(3\right)} \xi_{10}^{\left(3\right)} \xi_{01}^{\left(3\right)} \xi_{00}^{\left(3\right)}                
-2 l_{10} \left(4\xi_{11}^{\left(3\right)} \left(l_{00} \xi_{01}^{\left(3\right)}-l_{01} \xi_{00}^{\left(3\right)}\right)+l_{11} \left(l_{01} l_{00}+4\xi_{01}^{\left(3\right)} \xi_{00}^{\left(3\right)}\right)\right)  \Big].~~~~~~~
\eea

Using Eq.~(\ref{eq:gammajn}) we find the $\gamma^{\left(j\right)}_{n}$'s required for the first-order and second-order shear coefficients,
\bea
\gamma^{\left(0\right)}_{n}&=&\frac{2}{5}\gamma_{n},    \\
\gamma_{n}^{\left(1\right)}&=&\frac{4}{5}\gamma_{n}=\gamma_{n}^{\left(3\right)}, \\
\gamma_{n}^{\left(2\right)}&=&\gamma_{n}^{\left(4\right)}=0, 
\eea
where,
\bea 
\gamma_{0}&=&10\frac{S_{3}^{-2}}{S_{2}^{-1}},\\
\gamma_{1}&=&10\left\{\left(\frac{7}{2}+z\right)\frac{S_{3}^{-2}}{S_{2}^{-1}}-z\left(\frac{6}{z}\frac{S_{3}^{-3}}{S_{2}^{-1}}+\frac{S_{2}^{-2}}{S_{2}^{-1}}\right)\right\}. 
\eea
Similarly using Eq.~(\ref{eq:appcnj})-(\ref{eq:applnj}) we get, 
\bea
  c_{nj}&=&\frac{2}{5}c'_{nj},  \\
  d_{nj}&=&l_{nj}= \frac{4}{5}c'_{nj}, 
\eea 
where the expression for $c'_{nj}$ are given in App.~(\ref{sec:formula}). The  following $ \xi^{\left(a\right)}_{nj}$'s (from Eq.~\eqref{eq:xi}) are used from  for the first-order and the second-order shear viscosities,
\bea
 \xi^{\left(2\right)}_{nj}&=&\xi^{\left(4\right)}_{nj}=0,\\
\xi^{\left(1\right)}_{00}&=&\xi^{\left(3\right)}_{00}=\frac{4}{5}\frac{qB}{\rho}\gamma_{0}, \\
\xi^{\left(1\right)}_{01}&=&\xi^{\left(1\right)}_{10}=\xi^{\left(3\right)}_{01}=\xi^{\left(3\right)}_{10}=\frac{4}{5}\frac{qB}{\rho}\gamma_{1}, \\
\xi^{\left(1\right)}_{11}&=&\xi^{\left(3\right)}_{11}=\frac{8qB}{\rho}\Bigg[\left\{\left(\frac{7}{2}+z\right)^{2}+z^{2}\right\}\frac{S_{3}^{-2}}{S_{2}^{-1}}-2z\left(\frac{7}{2}+z\right)   \nonumber\\
& & ~~~~~\times\left(\frac{6}{z}\frac{S_{3}^{-3}}{S_{2}^{-1}}+\frac{S_{2}^{-2}}{S_{2}^{-1}}\right)         
+7z\left(\frac{6}{z}\frac{S_{3}^{-4}}{S_{2}^{-1}}+\frac{S_{2}^{-3}}{S_{2}^{-1}}\right)\Bigg].
\eea

\section{Detail Derivation of the Bulk Viscosity}\label{Appendix_Bulk}

The order-by-order calculation of the bulk viscosity is similar to the shear viscosity. The kinetic theory definition of the bulk viscous stress is
given by,
\bea
\Pi\triangle^{\mu\nu}= \zeta^{\eta\delta} \left(\partial_{\eta}U_{\delta}\right)\triangle^{\mu\nu}=\int d\Gamma_{p}p^{\mu}p^{\nu}f^{0}\left(1+A_{0}f^{0}\right) \phi.
\eea  
Now using the decomposition $\zeta^{\eta\delta}= \zeta_{||}C_{\left(0\right)}^{\eta\delta}+\zeta_{\perp}C_{\left(1\right)}^{\eta\delta}+\zeta_{\times}C_{\left(2\right)}^{\eta\delta}$ we have 
\bea
\left(\zeta_{||}C_{\left(0\right)}^{\eta\delta}+\zeta_{\perp}C_{\left(1\right)}^{\eta\delta}+\zeta_{\times}C_{\left(0\right)}^{\eta\delta}\right)\left(\partial_{\eta}U_{\delta}\right)\triangle^{\mu\nu}=\int d\Gamma_{p}p^{\mu}p^{\nu}f^{0}\left(1+A_{0}f^{0}\right)\left[\sum_{n=0}^{2}\mathcal{Z}_{n}C_{\left(n\right)}^{\eta\delta}\right]\left(\partial_{\eta}U_{\delta}\right), \nonumber \\  \label{Ap-Bulk1}
\eea 
where $C^{\eta\delta}_{(0,1,2)}$ are given in Eqs. \eqref{eq:C0app},\eqref{eq:C1app},\eqref{eq:C2app}.
To calculate the $ \zeta_{||}, \zeta_{\perp},\zeta_{\times}$ we contract both sides of the Eq.(\ref{Ap-Bulk1}) with $\triangle_{\mu\nu}P^{n}_{\eta\delta}$ and get,
\bea
\zeta_{||}=\frac{1}{3}\int d\Gamma_{p} f^{0}\left(1+A_{0}f^{0}\right)\triangle_{\mu\nu}p^{\mu}p^{\nu}\mathcal{Z}_{0}=\int d\Gamma_{p} f^{0}\left(1+A_{0}f^{0}\right)Q\mathcal{Z}_{0},\\
\zeta_{\perp}=\frac{1}{3}\int d\Gamma_{p} f^{0}\left(1+A_{0}f^{0}\right)\triangle_{\mu\nu}p^{\mu}p^{\nu}\mathcal{Z}_{1}=\int d\Gamma_{p} f^{0}\left(1+A_{0}f^{0}\right)Q\mathcal{Z}_{1},\\
\zeta_{\times}=\frac{1}{3}\int d\Gamma_{p} f^{0}\left(1+A_{0}f^{0}\right)\triangle_{\mu\nu}p^{\mu}p^{\nu}\mathcal{Z}_{2}=\int d\Gamma_{p} f^{0}\left(1+A_{0}f^{0}\right)Q\mathcal{Z}_{2}.
\eea 

These are the formulas reported in the main text Eqs.~\eqref{eq:zetapar},\eqref{eq:zetaperp},\eqref{eq:zetacross}. In the next-step we need to evaluate 
the unknowns $\mathcal{Z}_{n}$'s, for that we consider the BUU equation for the case of only non-zero  $\nabla_{\alpha}U^{\alpha}$ we get:
\bea
\frac{f^{0}\left(1+A_{0}f^{0}\right)}{T}Q\nabla_{\sigma}U^{\sigma}=-\sum_{n=0}^{2}\int d\Gamma_{k}d\Gamma_{p'}d\Gamma_{k'}\Big[f^{0}_{p}f^{0}_{k}\left(1+A_{0}f^{0}_{p'}\right)\left(1+A_{0}f^{0}_{k'}\right)\nonumber\\
\times \left(\mathcal{Z}_{n}^{p}+\mathcal{Z}_{n}^{k}-\mathcal{Z}_{n}^{p'}-\mathcal{Z}_{n}^{k'}\right)WC_{\left(n\right)\eta\delta}\partial^{\eta}U^{\delta}\Big]~\nonumber\\
-qF^{\mu\nu}p_{\nu}f^{0}\left(1+A_{0}f^{0}\right)\frac{\partial}{\partial p^{\mu}}\left(\sum_{n=0}^{2}\mathcal{Z}_{n}C_{\left(n\right)\eta\delta}\right)\partial^{\eta}U^{\delta}.
\eea 
The last term in the above equation vanishes since,
\bea
qF^{\mu\nu}p_{\nu}\frac{\partial}{\partial p^{\mu}}\left(\sum_{n=0}^{2}\mathcal{Z}_{n}C_{\left(n\right)\eta\delta}\right)=qF^{\mu\nu}p_{\nu}\left(\sum_{n=0}^{2}\frac{\partial \mathcal{Z}_{n}}{\partial \tau}U_{\mu}C_{\left(n\right)\eta\delta}\right)=0.
\eea 
Thus we have,
\bea
\frac{f^{0}\left(1+A_{0}f^{0}\right)}{T}Q\triangle_{\eta\delta}\left(\partial^{\eta}U^{\delta}\right)
&=&-\sum_{n=0}^{2}\int d\Gamma_{k}d\Gamma_{p'}d\Gamma_{k'}\Big[f^{0}_{p}f^{0}_{k}\left(1+A_{0}f^{0}_{p'}\right)\left(1+A_{0}f^{0}_{k'}\right)\nonumber \\
&\times  &\left(\mathcal{Z}_{n}^{p}+\mathcal{Z}_{n}^{k}-\mathcal{Z}_{n}^{p'}-\mathcal{Z}_{n}^{k'}\right) WC_{\left(n\right)\eta\delta}\left(\partial^{\eta}U^{\delta}\right)\Big].
\eea 
Using the relation $\triangle_{\mu\nu}=P^{0}_{\mu\nu}+P^{1}_{\mu\nu}+P^{-1}_{\mu\nu}$ in the above equation and contracting both sides with $P^{n\eta\delta}$ we get  we get the following three equations
\bea
\nonumber
\frac{f^{0}\left(1+f^{0}\right)}{T}Q&=&-\int d\Gamma_{k}d\Gamma_{p'}d\Gamma_{k'}f^{0}_{p}f^{0}_{k}\left(1+A_{0}f^{0}_{p'}\right)\left(1+A_{0}f^{0}_{k'}\right)
\left(\mathcal{Z}_{0}^{p}+\mathcal{Z}_{0}^{k}-\mathcal{Z}_{0}^{p'}-\mathcal{Z}_{0}^{k'}\right)W, \label{A1}\\
\eea
\bea
 \frac{f^{0}\left(1+f^{0}\right)}{T}Q&=&-\int d\Gamma_{k}d\Gamma_{p'}d\Gamma_{k'}f^{0}_{p}f^{0}_{k}\left(1+A_{0}f^{0}_{p'}\right)\left(1+A_{0}f^{0}_{k'}\right) \left(\mathcal{Z}_{1}^{p}+\mathcal{Z}_{1}^{k}-\mathcal{Z}_{1}^{p'}-\mathcal{Z}_{1}^{k'}\right)W\nonumber \\
& &-i\int d\Gamma_{k}d\Gamma_{p'}d\Gamma_{k'}f^{0}_{p}f^{0}_{k}\left(1+A_{0}f^{0}_{p'}\right)\left(1+A_{0}f^{0}_{k'}\right)\left(\mathcal{Z}_{2}^{p}+\mathcal{Z}_{2}^{k}-\mathcal{Z}_{2}^{p'}-\mathcal{Z}_{2}^{k'}\right)W,  \nonumber \\
\eea
\bea
\frac{f^{0}\left(1+f^{0}\right)}{T}Q&=&-\int d\Gamma_{k}d\Gamma_{p'}d\Gamma_{k'}f^{0}_{p}f^{0}_{k}\left(1+A_{0}f^{0}_{p'}\right)\left(1+A_{0}f^{0}_{k'}\right)\left(\mathcal{Z}_{1}^{p}+\mathcal{Z}_{1}^{k}-\mathcal{Z}_{1}^{p'}-\mathcal{Z}_{1}^{k'}\right)W \nonumber\\
& &+i\int d\Gamma_{k}d\Gamma_{p'}d\Gamma_{k'}f^{0}_{p}f^{0}_{k}\left(1+A_{0}f^{0}_{p'}\right)\left(1+A_{0}f^{0}_{k'}\right)\left(\mathcal{Z}_{2}^{p}+\mathcal{Z}_{2}^{k}-\mathcal{Z}_{2}^{p'}-\mathcal{Z}_{2}^{k'}\right)W \nonumber .\\ 
\eea 
From the above set of equations it is evident that the term associated with $\mathcal{Z}_2$ must vanish. Now integrating Eq.~(\ref{A1}) with weight $L_{j}^{\frac{1}{2}}\left(\tau\right)$ we have,
\bea
\int d\Gamma_{p}\frac{f^{0}\left(1+A_{0}f^{0}\right)}{T}QL_{j}^{\frac{1}{2}}\left(\tau'\right)&=&-\int d\Gamma_{p} d\Gamma_{k}d\Gamma_{p'}d\Gamma_{k'}f^{0}_{p}f^{0}_{k}\left(1+A_{0}f^{0}_{p'}\right)\left(1+A_{0}f^{0}_{k'}\right) \nonumber \\
&\times  &     \left(\mathcal{Z}_{0}^{p}+\mathcal{Z}_{0}^{k}-\mathcal{Z}_{0}^{p'}-\mathcal{Z}_{0}^{k'}\right)WL_{j}^{\frac{1}{2}}\left(\tau\right)   \nonumber\\
&=&-\frac{1}{4}\int d\Gamma_{p} d\Gamma_{k}d\Gamma_{p'}d\Gamma_{k'}f^{0}_{p}f^{0}_{k} \nonumber\\
&\times & \left(1+A_{0}f^{0}_{p'}\right)\left(1+A_{0}f^{0}_{k'}\right)W \left(\mathcal{Z}_{0}^{p}+\mathcal{Z}_{0}^{k}-\mathcal{Z}_{0}^{p'}-\mathcal{Z}_{0}^{k'}\right) \nonumber \\ 
&\times& \left(L^{\frac{1}{2}}_{j}\left(\tau'_{p}\right)+L^{\frac{1}{2}}_{j}\left(\tau'_{k}\right)-L^{\frac{1}{2}}_{j}\left(\tau'_{p'}\right)-L^{\frac{1}{2}}_{j}\left(\tau'_{k'}\right)\right) \nonumber \\ 
\eea 
The above equation can be rewritten as,
\bea
\frac{m}{\rho}\alpha_{j}=\left[\mathcal{Z}_{0},L_{j}^{\frac{1}{2}}\left(\tau'\right)\right],
\eea 
where,
\bea
\alpha_{j}&=&-\frac{m}{\rho T}\int d\Gamma_{p}f^{0}\left(1+A_{0}f^{0}\right)QL_{j}^{\frac{1}{2}}\left(\tau'\right),\\
\left[\mathcal{Z}_{0},L_{j}^{\frac{1}{2}}\left(\tau'\right)\right]&=&-\frac{m^{2}}{4\rho^{2}}\int d\Gamma_{p} d\Gamma_{k}d\Gamma_{p'}d\Gamma_{k'}f^{0}_{p}f^{0}_{k}\left(1+A_{0}f^{0}_{p'}\right)\left(1+A_{0}f^{0}_{k'}\right)WL_{j}^{\frac{1}{2}}\left(\tau'\right)~~~~~~ \label{alpha_j}\nonumber\\
\label{eq:C15}
&\times &\left(\mathcal{Z}_{0}^{p}+\mathcal{Z}_{0}^{k}-\mathcal{Z}_{0}^{p'}-\mathcal{Z}_{0}^{k'}\right)\left(L^{\frac{1}{2}}_{0}\left(\tau'_{p}\right)+L^{\frac{1}{2}}_{0}\left(\tau'_{k}\right)-L^{\frac{1}{2}}_{0}\left(\tau'_{p'}\right)-L^{\frac{1}{2}}_{0}\left(\tau'_{k'}\right)\right).  ~~~~~~~~
\eea 
Now using $\mathcal{Z}_{0}\left(B,\tau'\right)=\sum_{i=0}^{\infty}a^{0}_{i}\left(B\right)L^{\frac{1}{2}}_{i}\left(\tau'\right)$ we have,
\bea
\label{eq:C16}
\sum_{n=0}^{\infty}a^{0}_{n}\left[L_{n}^{\frac{1}{2}},L_{j}^{\frac{1}{2}}\right]=\sum_{n=0}^{\infty}a^{0}_{n}a_{nj}=\frac{m}{\rho}\alpha_{j}.
\eea 
Using the following properties: $L_{0}^{\frac{1}{2}}\left(\tau\right)=1$ and $L_{0}^{\frac{1}{2}}\left(\tau\right)=\frac{3}{2}-\tau$, $\int d\Gamma_{p}f^{0}\left(1+f^{0}\right)Q\phi=0$, one can show  $\alpha_{0}=\alpha_{1}=0$ and $a_{0j}=a_{j0}=a_{1j}=a_{j1}=0$.
If we limit the series to a finite $l$,
\bea
\sum_{n=0}^{l+1}a_{0}^{n\left(l\right)}a_{nj}=\frac{m}{\rho}\alpha_{j}.
\eea 
Finally truncating the series upto first-order we arrive at the expressions for $\left[\zeta_{n}\right]_1$s:
\bea
\left[\zeta_{||}\right]_{1}&=&\left[\zeta_{\perp}\right]_{1}=\frac{\rho T}{m}a_{0}^{2\left(1\right)}\alpha_{2}=T\frac{\alpha^{2}}{a_{22}},  \\
\left[\zeta_{\times}\right]_1&=&0.
\eea 
Similarly the second-order result is calculated and they are ,
\bea
\left[\zeta_{\times}\right]_2&=&0, \\
\left[\zeta_{||}\right]_{2}&=& \left[\zeta_{\perp}\right]_{2}=T\frac{\left(a_{33}\alpha_{2}^{2}-2\alpha_{2}\alpha_{3}a_{23}+a_{22}\alpha_{3}^{2}\right)}{\left(a_{22}a_{33}-a_{23}a_{32}\right)}.
\eea

\section{Detail Derivation of the Thermal Conductivity} \label{Appendix_Thermal}

In this case the $\phi$ in $\delta f $ has the following form:
\bea
\phi=\sum_{n=0}^{2}\mathcal{Y}_{n}C_{\left(n\right)}^{\alpha\beta}p_{\alpha}\left(T^{-1}\partial_{\beta}T-DU_{\beta}\right).
\eea 
The expression for the reduced heat flow is gven by,
\bea
I^{\mu}_{q}=\triangle^{\mu}_{\sigma}\left(U_{\nu}T^{\nu\sigma}-hN^{\sigma}\right),
\eea
where $h$ is the enthalpy per particle of the system. Using $\phi$ in the kinetic theory definition we have 
\bea
I^{\mu}_{q}=-\lambda^{\mu\nu}\triangle^{\beta}_{\nu}\left(\partial_{\beta}T-TDU_{\beta}\right)=\int d\Gamma_{p}~p^{\sigma}\triangle^{\mu}_{\sigma}f^{0}\left(1+A_{0}f^{0}\right)\phi_{p}\left(p\cdot U-h\right).
\eea 
Like for the other cases, the thermal conductivity tensor can be decomposed as $\lambda^{\mu\nu}=\sum_{n=0}^{2}\lambda_{n}C_{\left(n\right)}^{\mu\nu}$;
hence we have
\bea
\label{eq:heatFlow}
\nonumber
-\sum_{n=0}^{2}\lambda_{n}C_{\left(n\right)}^{\mu\nu}\triangle^{\beta}_{\nu}\left(\partial_{\beta}T-TDU_{\beta}\right)  &=&   \int d\Gamma_{p}~p^{\sigma}\triangle^{\mu}_{\sigma}f^{0}\left(1+A_{0}f^{0}\right)\left(p\cdot U-h\right)\\
&\times &\sum_{n=0}^{2}\mathcal{Y}_{n}C_{\left(n\right)}^{\alpha\beta}p_{\alpha}\left(T^{-1}\partial_{\beta}T-DU_{\beta}\right).
\eea
As usual, we obtain $\lambda_0$ by contracting both sides of the above equation with $P_{\left(0\right)}^{\beta\theta}$, 
\bea 
\lambda_{0}= \frac{1}{T} \int d\Gamma_{p}~p^{\sigma}\triangle^{\mu}_{\sigma}p^{\alpha}f^{0} \left(1+A_{0}f^{0}\right)\left(p\cdot U-h\right) \mathcal{Y}_{0}P_{\left(0\right)\alpha\mu}.  \label{lambda0}
\eea 
Contracting Eq.~(\ref{eq:heatFlow}) with  $P_{\left(1\right)}^{\beta\theta}$, and $P_{\left(2\right)}^{\beta\theta}$ gives
\bea 
\lambda_{1}+i\lambda_{2}&=&  \frac{1}{T}\int d\Gamma_{p}~p^{\sigma}\triangle^{\mu}_{\sigma}p^{\alpha}f^{0} \left(1+A_{0}f^{0}\right)\left(\mathcal{Y}_{1}+i\mathcal{Y}_{2}\right)\left(p\cdot U-h\right)P_{\left(1\right)\alpha\mu}, \\
\lambda_{1}-i\lambda_{2}&=&  \frac{1}{T}  \int d\Gamma_{p}~p^{\sigma}\triangle^{\mu}_{\sigma}p^{\alpha}f^{0} \left(1+A_{0}f^{0}\right)\left(\mathcal{Y}_{1}-i\mathcal{Y}_{2}\right)  \left(p\cdot U-h\right) P_{\left(-1\right)\alpha\mu}.
\eea 
Solving the above two equations for $\lambda_1$ and $\lambda_2$ we have
\bea
\lambda_{1}=\frac{1}{2T}\int d\Gamma_{p}~p^{\sigma}\triangle^{\mu}_{\sigma}p^{\alpha}f^{0} \left(1+A_{0}f^{0}\right)\mathcal{Y}_{1} \left(p\cdot U-h\right) C_{\left(1\right)\alpha\mu}          ~~~~~~~~~~~\nonumber\\
+\frac{1}{2T}\int d\Gamma_{p}~p^{\sigma}\triangle^{\mu}_{\sigma}p^{\alpha}f^{0} \left(1+A_{0}f^{0}\right)\mathcal{Y}_{2}  \left(p\cdot U-h\right) C_{\left(2\right)\alpha\mu},  \label{lambda1}   \\
\lambda_{2}=\frac{1}{2T}\int d\Gamma_{p}~p^{\sigma}\triangle^{\mu}_{\sigma}p^{\alpha}f^{0} \left(1+A_{0}f^{0}\right)\mathcal{Y}_{2} \left(p\cdot U-h\right) C_{\left(1\right)\alpha\mu}       ~~~~~~~~~~~\nonumber\\ 
-  \frac{1}{2T}\int d\Gamma_{p}~p^{\sigma}\triangle^{\mu}_{\sigma}p^{\alpha}f^{0} \left(1+A_{0}f^{0}\right)\mathcal{Y}_{1}  \left(p\cdot U-h\right) C_{\left(2\right)\alpha\mu}.~~~ \label{lambda2}
\eea 
To derive the final expressions for $\lambda$'s as given in Eqs.~\eqref{eq:lamb0},\eqref{eq:lamb1},\eqref{eq:lamb2} we note the fact that the following relation holds for the integral appeared in the above equation 
\bea
\int d\Gamma_{p}~p^{\sigma}p^{\alpha}f^{0} \left(1+A_{0}f^{0}\right)\mathcal{Y}_{i}\left(p\cdot U-h\right)=I_{1}U^{\sigma}U^{\alpha}+I_{2}\triangle^{\sigma\alpha}.
\eea 
The first term disappears on contraction with $C_{\left(2\right)\alpha\mu}\triangle^{\mu}_{\sigma}$, and $I_{2}$ is 
\bea
I_{2}&=&-\frac{1}{3}\int d\Gamma_{p}~\vec{p}^{~2}f^{0} \left(1+A_{0}f^{0}\right)\mathcal{Y}_{i}\left(p\cdot U-h\right).
\eea 
Now we observe that $C_{\left(2\right)\alpha\mu}\triangle^{\mu\alpha}=0$ and $C_{\left(1\right)\alpha\mu}\triangle^{\mu\alpha}=-2$; hence
we get the final expressions Eqs.~\eqref{eq:lamb0},\eqref{eq:lamb1},\eqref{eq:lamb2}.
Further, we expand the coefficients $\mathcal{Y}_{j}$ as 
\bea
\label{eq:Yi}
\mathcal{Y}_{j}=\sum_{n=0}^{\infty}b_{n}^{\left(j\right)}L_{n}^{\frac{3}{2}}\left(\tau'\right).      
\eea 
Substituting the above expansion in Eq.\eqref{eq:lamb0} we get,
\bea
\label{eq:lambda0}
\lambda_{0}=-\frac{\rho T}{m}\sum_{n=0}^{\infty}b_{n}^{\left(0\right)}\beta^{\left(0\right)}_{n}, \label{lambda01}
\eea
where the expression for $\beta^{\left(0\right)}_{n}$ is given in Eq.~\eqref{eq:betatwoone} below. 
Substituting Eq.~\eqref{eq:Yi} in Eq.(\ref{lambda1}) and Eq.(\ref{lambda2}) we get,
\bea
\lambda_{1}=-\frac{\rho T}{2m}\sum_{n=0}^{\infty}\left(b^{\left(1\right)}_{n}\beta^{\left(1\right)}_{n}+b^{\left(2\right)}_{n}\beta^{\left(2\right)}_{n}\right),          \label{eq:lambda11}\\
\lambda_{2}=-\frac{\rho T}{2m}\sum_{n=0}^{\infty}\left(b^{\left(2\right)}_{n}\beta^{\left(1\right)}_{n}-b^{\left(1\right)}_{n}\beta^{\left(2\right)}_{n}\right).           \label{eq:lambda21}
\eea 
To derive the unknown coefficients $b^{\left(1\right)}_{n},b^{\left(2\right)}_{n}$ we use the BUU equation,
\bea
p^{\mu}\partial_{\mu}f^{0}&=&\frac{f^{0}\left(1+A_{0}f^{0}\right)}{T}\left(p\cdot U-h\right)p^{\mu}\triangle_{\mu}^{\sigma}T^{-1}\left(\partial_{\sigma}T-TDU_{\sigma}\right)     \nonumber\\
&=&-\int d\Gamma_{k}d\Gamma_{p'}d\Gamma_{k'}~f^{0}_{p}f^{0}_{k}\left(1+A_{0}f^{0}_{p'}\right)\left(1+A_{0}f^{0}_{k'}\right)\left(\phi_{p}+\phi_{k}-\phi_{p'}-\phi_{k'}\right)W         \nonumber\\
& &-qF^{\mu\nu}p_{\nu}f^{0}\left(1+A_{0}f^{0}\right)\frac{\partial \phi_{p}}{\partial p^{\mu}}.
\eea 
Now, the last term on the right hand side of the above equation can be further simplified as,
\bea
qF^{\mu\nu}p_{\nu}f^{0}\left(1+A_{0}f^{0}\right)\frac{\partial \phi_{p}}{\partial p^{\mu}}&=&qF^{\mu\nu}p_{\nu}f^{0}\left(1+A_{0}f^{0}\right)\sum_{n=0}^{2}\mathcal{Y}_{n}C_{\left(n\right)\alpha}^{~~~~\beta}\delta^{\alpha}_{\mu}T^{-1}\left(\partial_{\beta}T-TDU_{\beta}\right)                    \nonumber\\
&=&qF_{\mu\nu}p^{\nu}f^{0}\left(1+A_{0}f^{0}\right)\sum_{n=0}^{2}\mathcal{Y}_{n}C_{\left(n\right)}^{\mu\beta}T^{-1}\left(\partial_{\beta}T-TDU_{\beta}\right)
\eea 
Using the techniques described before for the shear and bulk viscosity we derive the $b^{\left(1\right)}_{n},b^{\left(2\right)}_{n}$ by 
multiplying both sides of  with the projection tensor $P_{\left(n\right)\mu\beta}$ .
For $P_{\left(0\right)\mu\beta}$ we have 
\bea
-\frac{f^{0}\left(1+A_{0}f^{0}\right)}{T}\left(p\cdot U-h\right)p^{\mu}P_{\left(0\right)\mu\beta}&=& 
-\int d\Gamma_{k}d\Gamma_{p'}d\Gamma_{k'}f^{0}_{p}f^{0}_{p}\left(1+A_{0}f^{0}_{p'}\right)\left(1+A_{0}f^{0}_{k'}\right)\nonumber\\ 
&\times  &\left[\mathcal{Y}_{0}^{p}p^{\mu}+\mathcal{Y}_{0}^{k}k^{\mu}-\mathcal{Y}_{0}^{p'}p'^{\mu}-\mathcal{Y}_{0}^{k'}k'^{\mu}\right]P_{\left(0\right)\mu\beta}W. ~~ ~~~~~~       \label{k01}
\eea    
Similarly for $P_{\left(1\right)\mu\beta}$ and $P_{\left(-1\right)\mu\beta}$ we have 
\bea                                                                      
-\frac{f^{0}\left(1+A_{0}f^{0}\right)}{T}\left(p\cdot U-h\right)p^{\mu}P_{\left(1\right)\mu\beta}&=& -\int d\Gamma_{k}d\Gamma_{p'}d\Gamma_{k'}\bigg\{f^{0}_{p}f^{0}_{p}\left(1+A_{0}f^{0}_{p'}\right)\left(1+A_{0}f^{0}_{k'}\right) 
\nonumber\\
&\times  &\Bigg[\left(\mathcal{Y}_{1}^{p}+i\mathcal{Y}_{2}^{p}\right)p^{\mu}+\left(\mathcal{Y}_{1}^{k}+i\mathcal{Y}_{2}^{k}\right)k^{\mu} \nonumber \\ 
& & -\left(\mathcal{Y}_{1}^{p'}+i\mathcal{Y}_{2}^{p'}\right)p'^{\mu}-\left(\mathcal{Y}_{1}^{k'}+i\mathcal{Y}_{2}^{k'}\right)k'^{\mu}\Bigg]P_{\left(1\right)\mu\beta}W\bigg\}                    \nonumber\\
& &-qBp^{\mu}f^{0}\left(1+A_{0}f^{0}\right)\left(\mathcal{Y}_{2}-i\mathcal{Y}_{1}\right)P_{\left(1\right)\mu\beta},                          \label{k11}
\eea 

\bea
 -\frac{f^{0}\left(1+A_{0}f^{0}\right)}{T}\left(p\cdot U-h\right)p^{\mu}P_{\left(-1\right)\mu\beta}&=&  -\int d\Gamma_{k}d\Gamma_{p'}d\Gamma_{k'}\bigg\{f^{0}_{p}f^{0}_{p}\left(1+A_{0}f^{0}_{p'}\right)\left(1+A_{0}f^{0}_{k'}\right)   \nonumber\\
&\times &\Bigg[\left(\mathcal{Y}_{1}^{p}-i\mathcal{Y}_{2}^{p}\right)p^{\mu}+\left(\mathcal{Y}_{1}^{k}-i\mathcal{Y}_{2}^{k}\right)k^{\mu} \nonumber \\
& &-\left(\mathcal{Y}_{1}^{p'}-i\mathcal{Y}_{2}^{p'}\right)p'^{\mu}-\left(\mathcal{Y}_{1}^{k'}-i\mathcal{Y}_{2}^{k'}\right)k'^{\mu}\Bigg]P_{\left(-1\right)\mu\beta}W\bigg\}                                              \nonumber\\
& &-qBp^{\mu}f^{0}\left(1+A_{0}f^{0}\right)\left(\mathcal{Y}_{2}+i\mathcal{Y}_{1}\right)P_{\left(-1\right)\mu\beta}.     \label{k21}
\eea 
Now multiplying $L_{n}^{\frac{3}{2}}\left(\tau'\right)p^{\beta}$ on both sides of Eqs.~(\ref{k01}),\eqref{k11},\eqref{k21} and integrating we have  Eqs.~\eqref{eq:beta0lam},\eqref{eq:beta1lam},\eqref{eq:beta2lam} respectively, where 

\bea
\beta^{\left(m\right)}_{n}&=&-\frac{m}{\rho T^{2}}\int d\Gamma_{p}f^{0}\left(1+A_{0}f^{0}\right)\left(p\cdot U-h\right)p^{\mu}p^{\beta}C_{\left(m\right)\mu\beta}L_{n}^{\frac{3}{2}}\left(\tau'\right),  \nonumber  \\
&=&\frac{C_{\left(m\right)\mu\beta}\triangle^{\mu\beta}}{3}\beta_{n}.
\label{eq:betatwoone}
\eea 
Here the $\beta_{n}$'s are function of $z$ and the full form an be found in Ref.~\cite{Davesne}. For $n=1,2$ we have the following expression for $\beta_{n}$ 
\bea
\label{eq:beta_n}
\beta_{1}&=&-3z^{2}\left[1+5z^{-1} \frac{ S^{-2}_{3} }{ S^{-1}_{2} }-\left(\frac{ S^{-2}_{2} }{ S^{-1}_{2} }\right)\right],    \\
\beta_{2}&=&\left(\frac{7}{2}+z\right)\beta_{1}+\frac{z^{3}}{2}\left[ 3\left(\frac{ S^{-1}_{3} }{ S^{-1}_{2} }+\frac{30}{z^{2}}\frac{ S^{-3}_{3} }{ S^{-1}_{2} }+\frac{5}{z}\frac{ S^{-2}_{2} }{ S^{-1}_{2} }\right)-3\frac{ S^{-1}_{3} }{ S^{-1}_{2} }\left(1+\frac{5}{z}\frac{ S^{-2}_{3} }{ S^{-1}_{2} }\right)\right].
\eea
From Eq.~\eqref{eq:betatwoone} we have $\beta^{\left(0\right)}_{n}=-\frac{\beta_{n}}{3}$, $\beta^{\left(1\right)}_{n}=-\frac{2 \beta_{n}}{3}$ and $\beta^{\left(2\right)}_{n}=0$.
Further, we need to define $b^{\left(i\right)}_{mn}$ and $\theta^{\left(i\right)}_{mn}$ used in Eqs.~\eqref{eq:beta0lam},\eqref{eq:beta1lam},\eqref{eq:beta2lam}.
$b^{\left(i\right)}_{mn}$ has the following form (coming from the right hand side of the BUU equation while integrating with $L_{n}^{\frac{3}{2}}\left(\tau'\right)p^{\beta}$ as weight function to derive Eqs.~\eqref{eq:beta0lam},\eqref{eq:beta1lam},\eqref{eq:beta2lam})

\bea
\left[L_{m}^{\frac{3}{2}}\left(\tau'\right)p^{\mu},L_{n}^{\frac{3}{2}}\left(\tau'\right)p^{\beta}\right]_{C_{\left(i\right)}}=mT~b^{\left(i\right)}_{mn},
\eea 
where 
\bea
\frac{\rho^{2}}{m^{2}}\left[L_{m}^{\frac{3}{2}}\left(\tau'\right)p^{\mu},L_{n}^{\frac{3}{2}}\left(\tau'\right)p^{\beta}\right]_{C_{\left(i\right)}}&=& 
\frac{m^{2}}{4\rho^{2}}\int  d\Gamma_{p} d\Gamma_{k}d\Gamma_{p'}d\Gamma_{k'}\bigg\{f^{0}_{p}f^{0}_{p}\left(1+A_{0}f^{0}_{p'}\right)\left(1+A_{0}f^{0}_{k'}\right) \nonumber\\
&\times&\Big[L_{m}^{\frac{3}{2}}\left(\tau'_{p}\right)p^{\mu}+L_{m}^{\frac{3}{2}}\left(\tau'_{k}\right)k^{\mu} -L_{m}^{\frac{3}{2}}\left(\tau'_{p'}\right)p'^{\mu}-L_{m}^{\frac{3}{2}}\left(\tau'_{k'}\right)k'^{\mu}\Big] \nonumber \\ 
&\times &
\Big[L_{n}^{\frac{3}{2}}\left(\tau'_{p}\right)p^{\beta}+L_{n}^{\frac{3}{2}}\left(\tau'_{k}\right)k^{\beta}  -L_{n}^{\frac{3}{2}}\left(\tau'_{p'}\right)p'^{\beta}-L_{n}^{\frac{3}{2}}\left(\tau'_{k'}\right)k'^{\beta}\Big] \nonumber \\
& \times &C_{\left(i\right)\mu\beta}W\bigg\}.      \label{eqn:lambda_colli}
\eea
For convenience we also define 
\bea
b^{\left(i\right)}_{mn}=\frac{C_{\left(i\right)\mu\beta}\triangle^{\mu\beta}}{3}b_{mn},
\eea 
from the above definition we have $b^{\left(0\right)}_{mn}=-\frac{1}{3}b_{mn}$,  $b^{\left(1\right)}_{mn}=-\frac{2}{3}b_{mn}$ and  $b^{\left(2\right)}_{mn}=0$. The expression of $b_{mn}$ for differnt values of $m$ and $n$ are given in App.~(\ref{sec:formula}). 
Similarly $\theta^{\left(i\right)}_{mn}$ is defined from the right hand side of BUU equation while integrating with the weight  $L_{n}^{\frac{3}{2}}\left(\tau'\right)p^{\beta}$: 
\bea
\theta^{\left(j\right)}_{mn}=\frac{qBm}{\rho T^{2}}\int d\Gamma_{p}~ f^{0}\left(1+A_{0}f^{0}\right)p^{\mu}p^{\beta}C_{\left(j\right)\mu\beta}L_{n}^{\frac{3}{2}}\left(\tau_{p}\right)L_{m}^{\frac{3}{2}}\left(\tau_{p}\right).
\eea 
From the above definition, for $n=1,2$ we have
\bea
\theta^{\left(1\right)}_{mn}&=&\frac{2qBm}{3\rho T^{2}}\int d\Gamma_{p}~ f^{0}\left(1+A_{0}f^{0}\right)\left(\vec{p}\right)^{2}L_{n}^{\frac{3}{2}}\left(\tau_{p}\right)L_{m}^{\frac{3}{2}}\left(\tau_{p}\right),      \\
\theta^{\left(2\right)}_{mn}&=&0.
\eea 
For the first and second-order calculation of thermal conductivity we only require the following $\theta$'s
\bea 
\theta^{\left(1\right)}_{11}&=&\frac{2qB}{T}\left[\left\{\left(\frac{5}{2}+z\right)^{2}+z^{2}\right\}-2z\left(\frac{5}{2}+z\right)\frac{S^{-1}_{3}}{S^{-1}_{2}}+5z\frac{S^{-2}_{3}}{S^{-1}_{2}}\right],     \\
\theta^{\left(1\right)}_{12}&=&\frac{2qB}{T}\bigg[ \left(\frac{5}{2}+z\right)\left(\frac{35}{8}+\frac{7}{2}z+2z^{2}\right) -\left(\frac{85}{8}+\frac{17}{2}z+2z^{2}\right)z\frac{S^{-1}_{3}}{S^{-1}_{2}} \nonumber\\
& &+\frac{15}{2} \left(\frac{5}{2}+z\right)z\frac{S^{-2}_{3}}{S^{-1}_{2}}-15z\frac{S^{-3}_{3}}{S^{-1}_{2}}-\frac{5}{2}z^{2}   \frac{S^{-2}_{2}}{S^{-1}_{2}}  \bigg],\\
\theta^{\left(1\right)}_{22}&=&\frac{2qB}{T}\Bigg[ \left( \left(\frac{35}{8}+\frac{7}{2}z+\frac{z^{2}}{2}\right)^{2}+\left(\frac{85}{8}+\frac{17}{2}z+\frac{3z^{2}}{2}\right)z^{2}+\frac{z^{4}}{4}\right)  \nonumber\\
& &-5z^{2}\left(\frac{5}{2}+z\right) \frac{S^{-2}_{2}}{S^{-1}_{2}}+\frac{35}{4}z^{2} \frac{S^{-3}_{2}}{S^{-1}_{2}}-\left(\frac{5}{2}+z\right)\left( 2z \left(\frac{35}{8}+\frac{7}{2}z+\frac{z^{2}}{2}\right) +z^{3}\right) \frac{S^{-1}_{3}}{S^{-1}_{2}}  \nonumber\\
& &+\left( 5z \left(\frac{85}{8}+\frac{17}{2}z+\frac{3z^{2}}{2}\right) +\frac{5z^{3}}{2}\right) \frac{S^{-2}_{3}}{S^{-1}_{2}}-30z\left(\frac{5}{2}+z\right) \frac{S^{-3}_{3}}{S^{-1}_{2}}+\frac{105}{2} z\frac{S^{-4}_{3}}{S^{-1}_{2}}\Bigg].    
\eea 
Now, using the above relations we are ready to derive the final expression for $\lambda_{0,1,2}$. The first-order results are obtained by keeping 
terms upto first-order in expansion Eq.~\eqref{eq:Yi} and substituting in Eqs.~\eqref{eq:lambda0},\eqref{eq:lambda11},\eqref{eq:lambda21} we have:
\bea
\left[\lambda_{0}\right]_{1}&=&-\frac{\rho T}{m}b^{\left(0\right)}_{1}\beta^{\left(0\right)}_{1}=\frac{T}{m}\frac{\left(\beta^{\left(0\right)}_{1}\right)^{2}}{b^{\left(0\right)}_{11}}, \\
\left[\lambda_{1}\right]_{1}&=&-\frac{\rho T}{2m}\left(b^{\left(1\right)}_{1}\beta^{\left(1\right)}_{1}+b^{\left(2\right)}_{1}\beta^{\left(2\right)}_{1}\right)=\frac{ T}{2m}\frac{\left(\beta^{\left(1\right)}_{1}\right)^{2}b^{\left(1\right)}_{11}}{\left(b^{\left(1\right)}_{11}\right)^{2}+\left(\theta^{\left(1\right)}_{11}\right)^{2}}, \\
\left[\lambda_{2}\right]_{1}&=&-\frac{\rho T}{2m}\left(b^{\left(2\right)}_{1}\beta^{\left(1\right)}_{1}-b^{\left(1\right)}_{1}\beta^{\left(2\right)}_{1}\right)=\frac{ T}{2m}\frac{\left(\beta^{\left(1\right)}_{1}\right)^{2}\theta^{\left(1\right)}_{11}}{\left(b^{\left(1\right)}_{11}\right)^{2}+\left(\theta^{\left(1\right)}_{11}\right)^{2}}.
\eea 
Similarly truncating the series up to second-order we have 
\bea
\left[\lambda_{0}\right]_{2}&=&-\frac{T}{3m}\frac{\left(\beta_{1}\right)^{2}b_{22}-2\beta_{1}\beta_{2}b_{12}+\left(\beta_{2}\right)^{2}b_{11}}{b_{11}}, \\
\left[\lambda_{1}\right]_{2}&=&-\frac{\rho T}{2m}\left(b^{\left(1\right)}_{1}\beta^{\left(1\right)}_{1}+b^{\left(1\right)}_{2}\beta^{\left(1\right)}_{2}\right),  \\
\left[\lambda_{2}\right]_{2}&=&-\frac{\rho T}{2m}\left(b^{\left(2\right)}_{1}\beta^{\left(1\right)}_{1}+b^{\left(2\right)}_{2}\beta^{\left(1\right)}_{2}\right).
\eea
Where the coefficients $b^{(m)}_{n}$ ($m=n=1,2$) can be found from 
\bea
-\frac{1}{\rho}\beta^{\left(1\right)}_{1}&=&  b^{\left(1\right)}_{1}\left(b^{\left(1\right)}_{11}\right)+b^{\left(1\right)}_{2}\left(b^{\left(1\right)}_{21}\right)+b^{\left(2\right)}_{1}\left(\theta^{\left(1\right)}_{11}\right)+b^{\left(2\right)}_{2}\left(\theta^{\left(1\right)}_{21}\right) \\
-\frac{1}{\rho}\beta^{\left(1\right)}_{2}&=&  b^{\left(1\right)}_{1}\left(b^{\left(1\right)}_{12}\right)+b^{\left(1\right)}_{2}\left(b^{\left(1\right)}_{22}\right)+b^{\left(2\right)}_{1}\left(\theta^{\left(1\right)}_{12}\right)+b^{\left(2\right)}_{2}\left(\theta^{\left(1\right)}_{22}\right)\\
0&=& b^{\left(1\right)}_{1}\left(\theta^{\left(1\right)}_{11}\right)+b^{\left(1\right)}_{2}\left(\theta^{\left(1\right)}_{21}\right)-b^{\left(2\right)}_{1}\left(b^{\left(1\right)}_{11}\right)-b^{\left(2\right)}_{2}\left(b^{\left(1\right)}_{21}\right)      \\
0&=&  b^{\left(1\right)}_{1}\left(\theta^{\left(1\right)}_{12}\right)+b^{\left(1\right)}_{2}\left(\theta^{\left(1\right)}_{22}\right)-b^{\left(2\right)}_{1}\left(b^{\left(1\right)}_{12}\right)-b^{\left(2\right)}_{2}\left(b^{\left(1\right)}_{22}\right).
\eea

\section{Other Useful Formulas} \label{sec:formula}
\label{app:Other}
For all the calculations presented in this manuscript we have further made use of the following simplification of the collision term.
Starting with  Eqs.~\eqref{eq:C15},\eqref{eq:C16} we transform the incoming and outgoing four-momentum of the colliding particles
$p^{\mu},k^{\mu},p^{\prime\mu},k^{\prime\mu}$ to the following relative and global momentum:
\bea
g_{\alpha}&=&\frac{1}{2} \left(k_{\alpha}-p_{\alpha}\right),    \nonumber\\
g'_{\alpha}&=&\frac{1}{2} \left(k^{\prime}_{\alpha}-p^{\prime}_{\alpha}\right),    \\
P_{\alpha}&=&p_{\alpha}+k_{\alpha}=p^{\prime}_{\alpha}+k^{\prime}_{\alpha}=P^{\prime}_{\alpha}.
\eea 
Further using the polar-coordinate and rapidity $\psi$ we express $P^{\alpha}=P(\cosh\psi,\sinh\psi\sin\bar{\theta}\cos\bar{\varphi}, \\
\sinh\psi\sin\bar{\theta}\sin\bar{\varphi},\sinh\psi\cos\bar{\theta} )$. For convenience we use the following orthonormal i.e.,$( e^{\alpha}_{\left(i\right)}e_{\alpha\left(j\right)}=\delta_{ij}$) space-like four vectors in subsequent calculations:
\bea 
e^{\alpha}_{\left(1\right)}&=&\left(0,\cos\bar{\theta}\cos\bar{\varphi},\cos\bar{\theta}\sin\bar{\varphi},-\sin\bar{\theta}\right), \nonumber\\
e^{\alpha}_{\left(2\right)}&=&\left(0,-\sin\bar{\varphi},\cos\bar{\varphi},0\right),   \\
e^{\alpha}_{\left(3\right)}&=&\left(\sinh\psi,\cosh\psi\sin\bar{\theta}\cos\bar{\varphi},\cosh\psi\sin\bar{\theta}\sin\bar{\varphi},\cosh\psi\cos\bar{\theta}\right) .      \nonumber
\eea 
$e^{\alpha}_{\left(i\right)}$ are also orthogonal to the $P^{\alpha}$, i.e., $ e^{\alpha}_{\left(i\right)}P^{\alpha}=0$ for ($i=1,2,3$). 
In the centre of mass frame, $P^{\alpha}$ reads $P^{\alpha}=\left(P,\bf{0} \right)$, and $g^{\alpha}=\left(0,\textbf{\textit{g}}\right)$, and $e^{\alpha}_{\left(i\right)}=\left(0,\textbf{\textit{e}}_{i}\right)$ have only space components. Therefore we can write any three vector $\textbf{\textit{g}}$ in terms of the basis vectors as:
\bea 
\textbf{\textit{g}}=g\left(\sin\theta\cos\varphi~\textbf{\textit{e}}_{1}+\sin\theta\sin\varphi~\textbf{\textit{e}}_{2}+\cos\theta~\textbf{\textit{e}}_{3}\right).
\eea 
Here $\theta,\phi$ are the angle subtended by $\textbf{\textit{g}}$ with the $\textbf{\textit{e}}_{i}$.
In the permanent local rest frame we can define a four-velocity $U^{\alpha}=(\cosh\chi,\sinh\chi \bar{\textbf{\textit{e}}}_{3})$ so that $P_{0}=P^{\alpha}U_{\alpha}$  (see Ref.~\cite{De.Groot,De.Groot2} for more details). 
Let us define $\Theta$ the scattering angle in the center of mass frame $\left(\textbf{\textit{g}}\cdot\textbf{\textit{g}}'=g^{2}\cos\Theta\right)$  and $\psi$ such as $\sinh\psi=g/mc$ so that $P=2mc\cosh\psi$.  
\bea 
P_{0}&=&2mc\cosh\psi \cosh\chi,     \nonumber\\
g_{0}&=&mc\sinh\psi\sinh\chi\cos\theta,  \nonumber\\
g'_{0}&=&mc\sinh\psi\sinh\chi\cos\theta',    
\eea 
 where $\cos\theta'=\cos\theta\cos\Theta-\sin\theta\sin\Theta\cos\phi $ .
 Using these transformed variables we can write:
 \bea
 f^{0}_{p} f_{k}^{0}\left(1+A_{0} f^{0}_{p^{\prime}}\right)\left(1+A_{0} f^{0}_{k^{\prime}}\right)&=&\frac{A_{0}^{-2} e^{\left(-2 \mu / T\right)} e^{2 z \cosh \psi \cosh \psi}}{(e^{E}-1)(e^{F}-1)(e^{G}-1)(e^{H}-1)}, \\
 \eea
 where 
\bea 
E&=&z\left(\cosh\psi\cosh\chi-\sinh\psi\sinh\chi\cos\theta \right)-\mu/T,         \\
F&=&z\left(\cosh\psi\cosh\chi-\sinh\psi\sinh\chi\cos\theta' \right)-\mu/T,        \\
G&=&E+2z\sinh\psi\sinh\chi\cos\theta,                    \\
H&=&F+2z\sinh\psi\sinh\chi\cos\theta' .        \eea    
Further, on evaluation of the terms $a_{mn},b_{mn},c_{mn}$ (for bulk, thermal conductivity, and shear respectively)  we get Ref.~\cite{Davesne}.
\bea 
a_{22}&=&A_{22},      \\
a_{23}&=&A_{23}+\left(\frac{7}{2}+z\right) A_{22},\\
a_{33}&=&A_{33}+2\left(\frac{7}{2}+z\right) A_{23}+\left(\frac{7}{2}+z\right)^{2} A_{22}.
\eea 
where
\bea 
A_{22}&=&\chi_{0420},\\
A_{23}&=&-z~\chi_{1420},\\
A_{33}&=&z^{2}\chi_{2420}.
\eea 
$\chi$'s are multidimensional integration defined later in this section. Similarly 
\bea 
b_{11}&=&B_{11}^{\prime},\\
b_{12}&=&B_{12}^{\prime}+\left(\frac{7}{2}+z\right) B_{11}^{\prime},\\
b_{22}&=&B_{22}^{\prime}+2\left(\frac{7}{2}+z\right) B_{12}^{\prime}+\left(\frac{7}{2}+z\right)^{2} B_{11}^{\prime}.
\eea 
Where
\bea 
B'_{11}&=&B_{11}-4\frac{A_{22}}{z^{2}},\\
B'_{12}&=&B_{12}-6\frac{A_{23}}{z^{2}},\\
B'_{22}&=&B_{22}-9\frac{A_{33}}{z^{2}}.
\eea 

\bea 
B_{11}&=&-4\chi'_{0402100},\\
B_{11}&=&--4z\chi'_{1412100},\\
B_{11}&=&=-4z^{2}\chi'_{2422100}+z^{3}\chi''''_{240421}. 
\eea 
And
\bea
c'_{00}&=&C_{00}^{\prime}\\
c'_{01}&=&C_{01}^{\prime}+\left(\frac{7}{2}+z\right) C_{00}^{\prime},\\
c'_{11}&=&C_{11}^{\prime}+2\left(\frac{7}{2}+z\right) C_{01}^{\prime}+\left(\frac{7}{2}+z\right)^{2} C_{00}^{\prime},
\eea 
where
\bea 
C'_{00}&=&C_{00}+2z^{-1}B_{11}+\frac{8}{3}\frac{A_{22}}{z^{2}}, \\
C'_{01}&=&C_{01}+4z^{-1}B_{12}+8\frac{A_{23}}{z^{2}},\\
C'_{11}&=&C_{11}+8z^{-1}B_{22}+24\frac{A_{33}}{z^{2}}.
\eea 
Finally $C_{mn}$'s are evaluated in term of the $\chi$ integrals (defined below):
\bea 
C_{00}&=&8\chi'''_{040002},\\
C_{01}&=&-8z\chi''''_{141002},\\
C_{11}&=&8z^{2}\left(\chi'''_{242002}-\chi''_{24022}\right).
\eea 
The relevant $\chi$ integrals are:
\bea 
\chi_{ijkl}&=&\frac{2z^{6}e^{-2\mu/T}}{\left(S^{-1}_{2}\right)^{2}}\int_{0}^{\infty}d\psi \left(\cosh\psi~\sinh\psi\right)^{3}\cosh^{i}\psi ~\sinh^{j}\psi     \nonumber\\
&\times & \int_{0}^{\pi}d\Theta~\sin\Theta~\sigma\left(\psi\Theta\right)\int_{0}^{\infty}d\chi~\sinh^{2}\chi~\cosh^{i}\chi~\sinh^{j}\chi       \nonumber \\
&\times & \int_{0}^{2\pi}d\phi \int_{0}^{\pi}d\theta~\sin\theta \frac{e^{2z\cosh\psi~\cosh\chi}\left[\cos^{2}\theta-\cos^{2}\theta'\right]^{k}\left[\cos^{2}\theta+\cos^{2}\theta'\right]^{l}}{\left(e^{E}-1\right)\left(e^{F}-1\right)\left(e^{G}-1\right)\left(e^{H}-1\right)},~~~~~~   \\
\chi'_{ijklmno}&=&\frac{2z^{5}e^{-2\mu/T}}{\left(S^{-1}_{2}\right)^{2}}\int_{0}^{\infty}d\psi \left(\cosh\psi~\sinh\psi\right)^{3}\cosh^{i}\psi ~\sinh^{j}\psi     \nonumber\\
&\times & \int_{0}^{\pi}d\Theta~\sin\Theta~\sigma\left(\psi\Theta\right)\int_{0}^{\infty}d\chi~\sinh^{2}\chi~\cosh^{k}\chi~\sinh^{l}\chi       \nonumber \\
&\times & \int_{0}^{2\pi}d\phi \int_{0}^{\pi}d\theta~\sin\theta \frac{e^{2z\cosh\psi~\cosh\chi}}{\left(e^{E}-1\right)\left(e^{F}-1\right)}\frac{\left[\cos^{2}\theta+\cos^{2}\theta'-2\cos\theta\cos\theta'\cos\Theta\right]^{m}}{\left(e^{G}-1\right)\left(e^{H}-1\right)}    \nonumber\\
&\times & \left[\cos^{4}\theta+\cos^{4}\theta'-\cos^{3}\theta\cos\theta'\cos\Theta-\cos\theta\cos^{3}\theta'\cos\Theta\right]^{n}     \nonumber\\
&\times &\left[\cos^{6}\theta+\cos^{6}\theta'-2\cos^{3}\theta\cos^{3}\theta'\cos\Theta\right]^{o},   \\
\chi''_{ijklm}&=&\frac{2z^{4}e^{-2\mu/T}}{\left(S^{-1}_{2}\right)^{2}}\int_{0}^{\infty}d\psi \left(\cosh\psi~\sinh\psi\right)^{3}\cosh^{i}\psi ~\sinh^{j}\psi     \nonumber\\
&\times & \int_{0}^{\pi}d\Theta~\sin\Theta~\sigma\left(\psi\Theta\right)\int_{0}^{\infty}d\chi~\sinh^{2}\chi~\cosh^{k}\chi~\sinh^{l}\chi       \nonumber \\
&\times & \int_{0}^{2\pi}d\phi \int_{0}^{\pi}d\theta~\sin\theta \frac{e^{2z\cosh\psi~\cosh\chi}}{\left(e^{E}-1\right)\left(e^{F}-1\right)}     \nonumber\\
&\times &\frac{\left[\cos^{m}\theta+\cos^{m}\theta'-2\cos^{m/2}\theta\cos^{m/2}\theta'\cos^{m/2}\Theta\right]^{m}}{\left(e^{G}-1\right)\left(e^{H}-1\right)},    \\
\chi'''_{ijklmn}&=&\frac{2z^{4}e^{-2\mu/T}}{\left(S^{-1}_{2}\right)^{2}}\int_{0}^{\infty}d\psi \left(\cosh\psi~\sinh\psi\right)^{3}\cosh^{i}\psi ~\sinh^{j}\psi     \nonumber\\
&\times & \int_{0}^{\pi}d\Theta~\sin\Theta~\sigma\left(\psi\Theta\right)\int_{0}^{\infty}d\chi~\sinh^{2}\chi~\cosh^{k}\chi~\sinh^{l}\chi       \nonumber \\
&\times & \int_{0}^{2\pi}d\phi \int_{0}^{\pi}d\theta~\sin\theta \frac{e^{2z\cosh\psi~\cosh\chi}}{\left(e^{E}-1\right)\left(e^{F}-1\right)}\frac{\left[\cos^{2}\theta+\cos^{2}\theta'\right]^{m}\left(1-cos^{n}\Theta\right)}{\left(e^{G}-1\right)\left(e^{H}-1\right)},   \\
\chi''''_{ijklmn}&=&\frac{2z^{4}e^{-2\mu/T}}{\left(S^{-1}_{2}\right)^{2}}\int_{0}^{\infty}d\psi \left(\cosh\psi~\sinh\psi\right)^{3}\cosh^{i}\psi ~\sinh^{j}\psi     \nonumber\\
&\times & \int_{0}^{\pi}d\Theta~\sin\Theta~\sigma\left(\psi\Theta\right)\int_{0}^{\infty}d\chi~\sinh^{2}\chi~\cosh^{k}\chi~\sinh^{l}\chi       \nonumber \\
&\times & \int_{0}^{2\pi}d\phi \int_{0}^{\pi}d\theta~\sin\theta \frac{e^{2z\cosh\psi~\cosh\chi}}{\left(e^{E}-1\right)\left(e^{F}-1\right)}\frac{\left[\cos^{2n}\theta+\left(-1\right)^{n}\cos^{2n}\theta'\right]^{m}}{\left(e^{G}-1\right)\left(e^{H}-1\right)}. 
\eea

\acknowledgments

	U.G. and V.R. would like to express their gratitude to the Department of Atomic Energy, India, for financial support. U.G. thanks Arghya Mukherjee 
	for his help in the numerical calculations. V.R. also acknowledges support from the Department of Science and Technology, India, through the INSPIRE faculty research grant (IFA-16-PH-167).



\end{document}